\documentclass[manuscript]{acmart}

\usepackage{framed}
\usepackage{xcolor}
\definecolor{framecolor}{rgb}{0.8,0.2,0.2} 
\renewenvironment{framed}{%
  \MakeFramed{\advance\hsize-\width \FrameRestore}}%
 {\endMakeFramed}

\usepackage{tabularx}

\AtBeginDocument{%
  }

\setcopyright{acmlicensed}
\copyrightyear{2018}
\acmYear{2018}
\acmDOI{XXXXXXX.XXXXXXX}
\acmConference[Conference acronym 'XX]{Make sure to enter the correct
  conference title from your rights confirmation email}{June 03--05,
  2018}{Woodstock, NY}

\acmISBN{978-1-4503-XXXX-X/2018/06}

\begin{document}

\title{"I Prompt, it Generates, we Negotiate."\\
Exploring Text-Image Intertextuality in Human–AI Co-Creation of Visual Narratives with VLMs}

\author{Mengyao Guo}
\email{guomengyao@hit.edu.cn}
\affiliation{%
  \institution{Harbin Institute of Technology, Shenzhen}
  \city{Shenzhen}
  \country{China}
}

\author{Kexin Nie}
\email{niekexinbella@gmail.com}
\authornote{Corresponding Author.}
\affiliation{%
  \institution{The University of Sydney}
  \city{Sydney}
  \country{Australia}}

\author{Ze Gao}
\email{zegaoap@hotmail.com}
\affiliation{%
  \institution{Hong Kong Polytechnic University}
  \city{Hong Kong SAR}
  \country{China}
}

\author{Black Sun}
\affiliation{%
 \institution{Aarhus University}
 \city{Aarhus}
 \country{Denmark}
 }
\email{blackthompson770@gmail.com}

\author{Xueyang Wang}
\email{niekexinbella@gmail.com}
\affiliation{%
  \institution{Institute for Network Sciences and Cyberspace, Tsinghua University}
  \city{Beijing}
  \country{China}}

\author{Jinda Han}
\email{jhan51@illinois.edu}
\affiliation{%
  \institution{University of Illinois at Urbana-Champaign}
  \city{Urbana}
  \country{United States}}

\author{Xingting Wu}
\email{xingtingwu@swin.edu.au}
\affiliation{%
  \institution{Swinburne University of Technology}
  \city{Melbourne}
  \country{Australia}}

\renewcommand{\shortauthors}{Guo et al.}

\begin{abstract}

Creating meaningful visual narratives through human-AI collaboration requires understanding how \textit{text-image intertextuality} emerges when textual intentions meet AI-generated visuals. We conducted a three-phase qualitative study with 15 participants using GPT-4o to investigate how novices navigate sequential visual narratives. Our findings show that users develop strategies to harness AI's semantic surplus by recognizing meaningful visual content beyond literal descriptions, iteratively refining prompts, and constructing narrative significance through complementary text-image relationships. We identified four distinct collaboration patterns and, through fsQCA's analysis, discovered three pathways to successful intertextual collaboration: Educational Collaborator, Technical Expert, and Visual Thinker. However, participants faced challenges, including cultural representation gaps, visual consistency issues, and difficulties translating narrative concepts into visual prompts. These findings contribute to HCI research by providing an empirical account of \textit{text-image intertextuality} in human-AI co-creation and proposing design implications for role-based AI assistants that better support iterative, human-led creative processes in visual storytelling.

\end{abstract}



\begin{teaserfigure}
\centering
  \includegraphics[width=\textwidth]{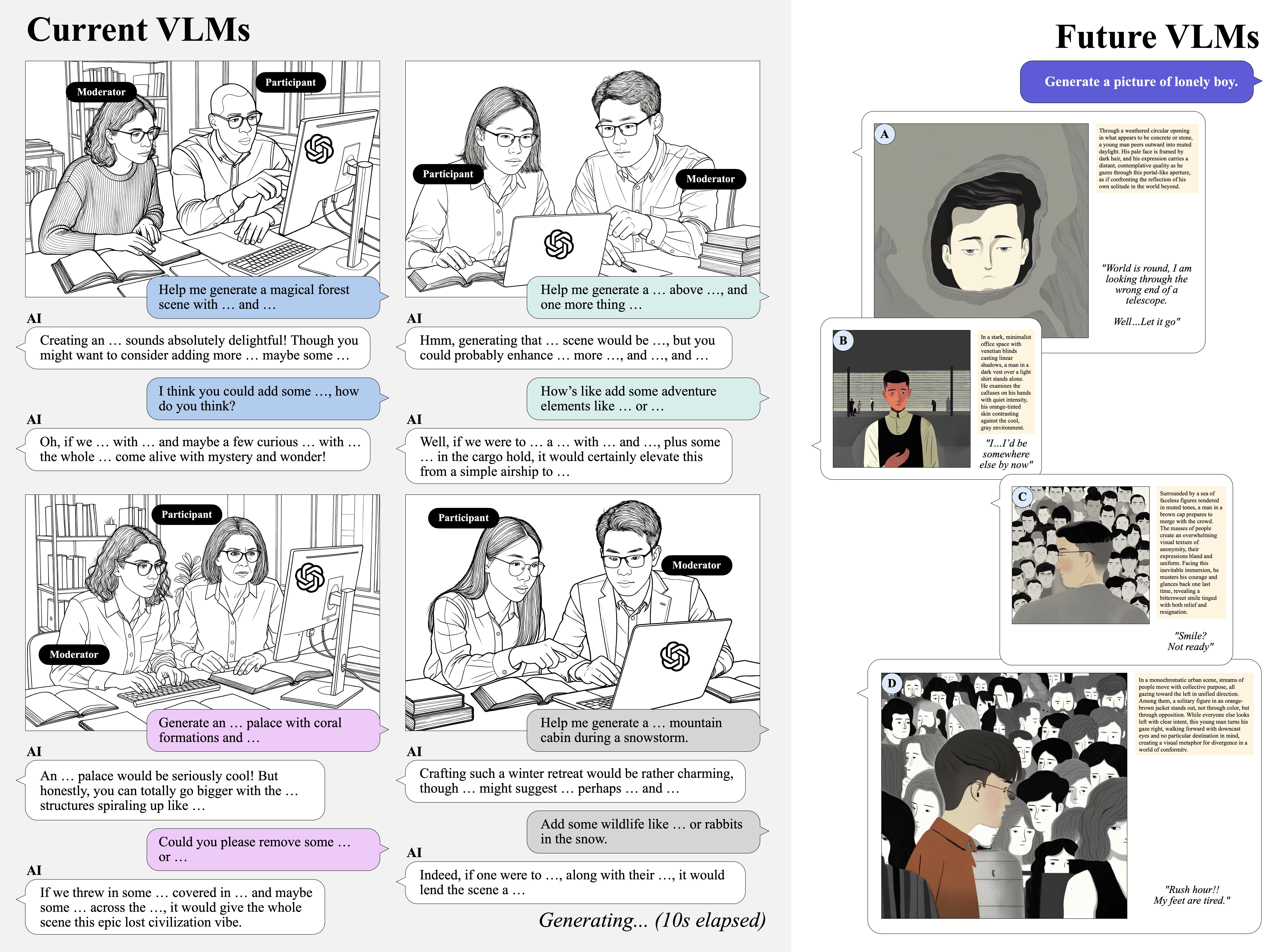}
  \caption{The comparison of current VLMs and future VLMs. Future VLM image generation ability, our envision of AI-driven text-image intertextuality. 1) The first panel creates resonance between the circular gaze and solitude text; 2) the second achieves perfect alignment between office alienation and professional confusion; 3) the third builds emotional tension between crowd pressure and resilient smiles; 4) the fourth reaches poetic harmony between walking against the flow and weary confession.}
  \Description{}
  \label{teaser}
\end{teaserfigure}

\maketitle

\section{Introduction}

\begin{figure*}
  \centering
  \includegraphics[width=\linewidth]{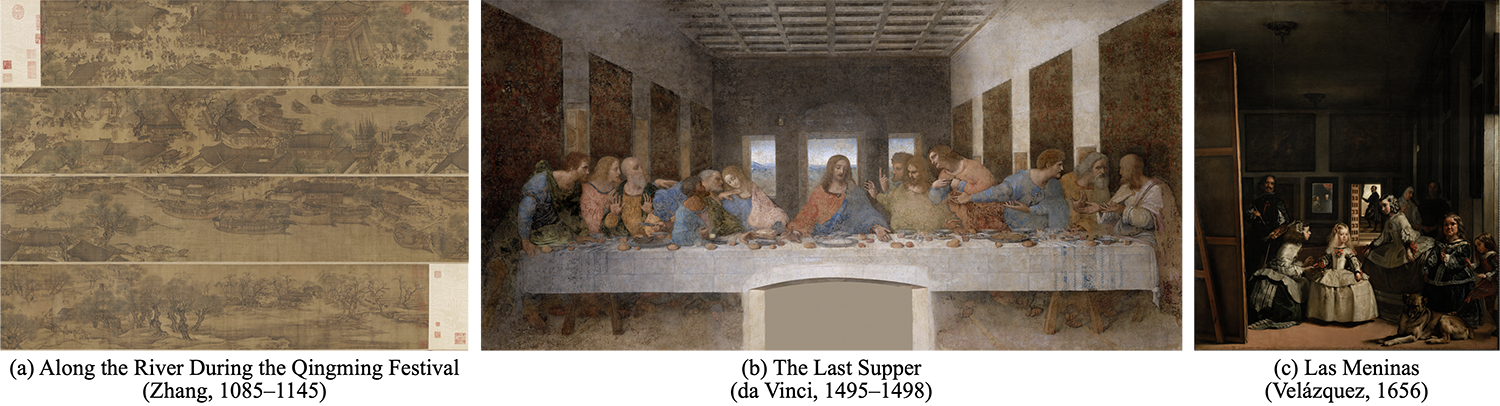}
  \caption{Throughout the communication process, cultural traditions have developed distinct approaches to visual representation. East Asian traditions often employ multiple perspectives and rhythmic elements to suggest narrative flow while maintaining holistic views~\cite{jing2024exploration,green2013rethinking}, as exemplified in "Along the River During the Qingming Festival" \cite{yu2023city}. In contrast, Western traditions have typically relied on linear perspective and the decisive moment\cite{sweet2008dialogue,panofsky2020perspective,cartier1993decisive}, as evident in works such as "The Last Supper" and "Las Meninas". These approaches reflect fundamentally different cognitive frameworks in the Eastern and Western worlds for organizing visual information rather than mere stylistic preferences. Differences in observation have yielded distinct models of visual narrative, underscoring the centrality of visual narrative to cultural representation and motivating our focus on text–image intertextuality.}
  \label{painting}
\end{figure*}

Visual narrative involves constructing meaning across sequential images and textual elements. From cave paintings and Chinese scrolls (see Fig.~\ref{painting}) to picture books, comics, and film, creators use composition, sequencing, symbolism, and color to convey events, context, and emotion. This visual method has been fundamental to human expression, predating written language~\cite{gombrich1995story}, and demonstrates how people transform the observed world into communicable images, symbols, and memories~\cite{arnheim2023visual}. Building on this tradition, contemporary narrative distributes storytelling functions across modalities~\cite{ellestrom2019transmedial}, with readers actively inferring events, affect, and causality between them~\cite{mar2011emotion}. In such settings, text and image are not direct translations of each other but work in concert, and meaning often emerges from their relationship rather than from either medium alone~\cite{long2021co}. We refer to this relation-centered meaning-making as text–image intertextuality (see Fig.~\ref{teaser}). 

With rapid advances in Vision Language Models (VLMs), non-experts can converse about images and synthesize visuals from natural language prompts within a single system~\cite{liu2024datasets}, such as GPT-5 and Nano Banana, or an AI agent to invoke models to achieve such text–image intertextuality. Unlike text-only Large Language Models (LLMs), VLMs integrate cross-modal understanding with text-conditioned image generation. When users provide concise prompts, outputs commonly include two components: \textit{Direct Mappings} that instantiate textual content, and \textit{Visual Augmentations} that add palettes, atmospheric cues, spatial compositions, or symbolic elements not specified in the text. While direct mappings establish the basic denotation of the prompt, it is the augmentations that often carry the connotative work of narrative, introducing mood, motif, and perspective that were not explicitly written. In sequential narratives, these augmentations frequently become the hinge points where panels cohere, where themes recur, and where irony or subtext emerges across frames. Our analysis focuses on how users perceive, evaluate, and negotiate these visual augmentations to construct text–image intertextuality, treating mappings as necessary scaffolding and augmentations as the primary sites of meaning-making. We consider that these augmentations can serve as loci of intertextual meaning by complementing, reframing, or extending textual intent.

Although GenAI for image generation and human–AI co-creation has advanced rapidly, in-depth accounts of the mechanisms that enable sequential visual narratives, especially text–image intertextuality, remain scarce~\cite{belz2024story,he2025leveraging,antony2025id}. Education and design studies frame storyboarding (one of the visual narrative forms) as a method for constructing stories and as a co-creative scaffold for sequencing, role-taking, and dramaturgy, extending into studio and classroom practices where students create concept videos and learn narrative structure~\cite{budach202213,read2022fostering,rasool2021transitioning,bjornstad2022anticipative,antoniou2021panel,thomas2021reflections}. Yet we still lack a clear portrait of how VLMs are used across an end-to-end narrative. To address this gap, we proposed \textbf{RQ1}, which examines current practice, including when and how participants invoked the VLM across stages, such as goal and idea, story structure and plot, dialogue, and full-text authoring, as well as how they divided labor between providing inputs and accepting or editing outputs.

In parallel, interactive agents and drawing systems support adjacent tasks such as character ideation and storyboard-oriented sketching, often using playful, abstract cues to stimulate visual thinking~\cite{ibarrola2023collaborative,jo2022interactive}. However, multimodal, co-creative evaluations situated within storyboarded workflows remain uncommon~\cite{antony2025id}, and integration into applied domains such as cultural heritage, where storyboard practices are diverse, remains emergent~\cite{nikolakopoulou2025making}. Moreover, limited analysis of text–image intertextuality constrains our understanding of how human textual inputs and AI-generated visuals relate and cohere across sequences. To address these gaps, we proposed \textbf{RQ2} on the extent to which users successfully establish intertextual meanings when using AI tools for visual narrative, and \textbf{RQ3} on the challenges that affect their ability to create these interconnected meanings, as well as on the future features and workflows users expect to better support text–image intertextuality and how these can be organised into functional roles for VLMs.

\textbf{RQ1:} How do users employ VLMs across the visual narrative workflow to allocate tasks and pursue text–image intertextuality?

\textbf{RQ2:} To what extent do users successfully establish text–image intertextual meanings when using VLMs for visual narrative?

\textbf{RQ3:} What challenges and desired features shape users' ability to create text–image intertextuality, including role-based workflows for VLMs?

In this paper, we conducted an in-depth, three-phase qualitative study with 15 participants to address these research questions. It consists of:~\textbf{Visual Narrative Concept Introduction},~\textbf{Think-Aloud Practice} with semi-open and open tasks, and a~\textbf{Reflective Interview}. We collected multimodal data to trace how users construct visual narratives with AI. For \textbf{RQ1}, participants utilised VLMs for rapid ideation and stylistic elaboration while maintaining control over narrative framing and continuity. They pursued intertextuality through four distinct collaboration approaches and learned to leverage AI's "semantic surplus," including visual augmentations that extend beyond literal descriptions. For \textbf{RQ2}, participants achieved moderate success in establishing \textit{text-image intertextuality} through externalised narrative structures and iterative refinement. Our fsQCA analysis identified three pathways to successful collaboration, with iteration willingness emerging as a critical factor. For \textbf{RQ3}, participants faced challenges including cultural representation gaps, visual consistency issues, and difficulties translating narrative to visual thinking. They desired role-based AI features such as continuity editors, motif trackers, and sequence-aware generation systems.

This work contributes to HCI and AI-mediated creativity research by:
\begin{itemize}
    \item Providing an empirical account of how non-experts negotiate meaning with VLMs in visual narrative workflows;
    \item Identifying key strategies and challenges in establishing \textit{text-image intertextuality};
    \item Proposing design implications for future AI tools, such as role-based assistants (e.g., continuity editor, motif tracker), that better support iterative, human-led creative processes.
\end{itemize}


\section{Background}

Visual narratives fundamentally operate through active cognition, not passive reception. Arnheim considered that \textit{"Visual perception is visual thinking"} \cite{arnheim2023visual}, suggesting that creative visual thinking is a cognitive skill that can be developed through practice. Schema theory also illuminates how viewers interpret visual narratives by activating archetypes based on prior experience~\cite{Jung2014}. These cognitive schemas help viewers fill gaps in visual sequences and infer causal relationships. As Bordwell noted: \textit{"The perceiver is not a passive receiver of data but an active mobilizer of structures and processes"} \cite{bordwell1991making}. Building on these foundations, Cohn et al.'s Grammar of Visual Narrative proposes that sequential images are processed hierarchically, with distinct narrative units (establishers, initials, peaks, releases) that advance the narrative~\cite{cohn2014grammar}.

\begin{figure*}
  \centering
  \includegraphics[width=\linewidth]{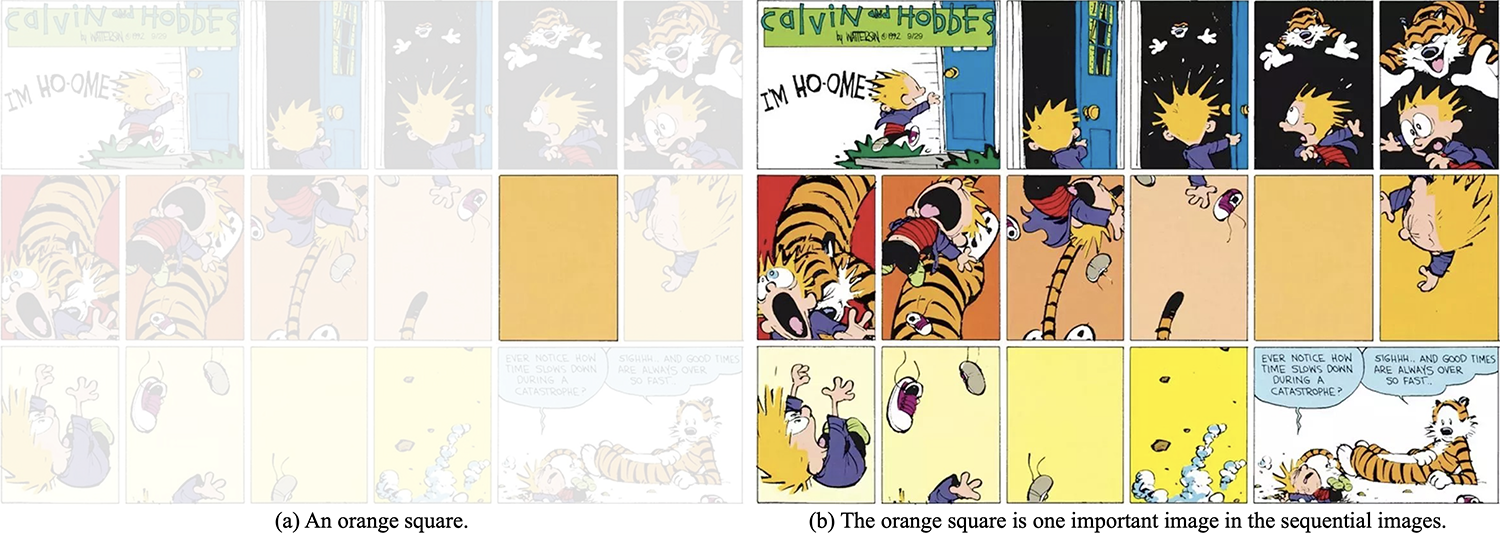}
  \caption{The orange square is one key component in the sequential relationship in \textit{Calvin and Hobbes}, Credit by Bill Watterson.}
  \label{orange}
\end{figure*}

Visual narrative coherence also depends on character consistency, environmental continuity, and clear cause-and-effect relationships that help viewers construct mental models~\cite{herman2009narrative}. These principles are realised through consistent character design, environmental anchoring, and visual motifs~\cite{nodelman2012picture,eisner2008graphic} (see Fig.~\ref{orange}). Unlike text or film, static visual narratives must suggest temporal relationships through spatial organisation~\cite{k2011temporality, pimenta2009representation}. Groensteen's concept of "arthrology" examines how panel relationships create meaning through sequential reading and distant "braiding" connections~\cite{groensteen2007system}. Similarly, Bang's principles of picture composition explain how elements like size, position, and contrast create narrative hierarchy and guide reading paths through visual compositions~\cite{driver1970herman}. These principles highlight the visual narrative's complexity, extending beyond textual illustration. The intricate panel relationships, spatial-temporal encoding, and multiple simultaneous information streams create unique challenges for human-AI collaborative narrative construction.

\begin{figure*}
  \centering
  \includegraphics[width=\linewidth]{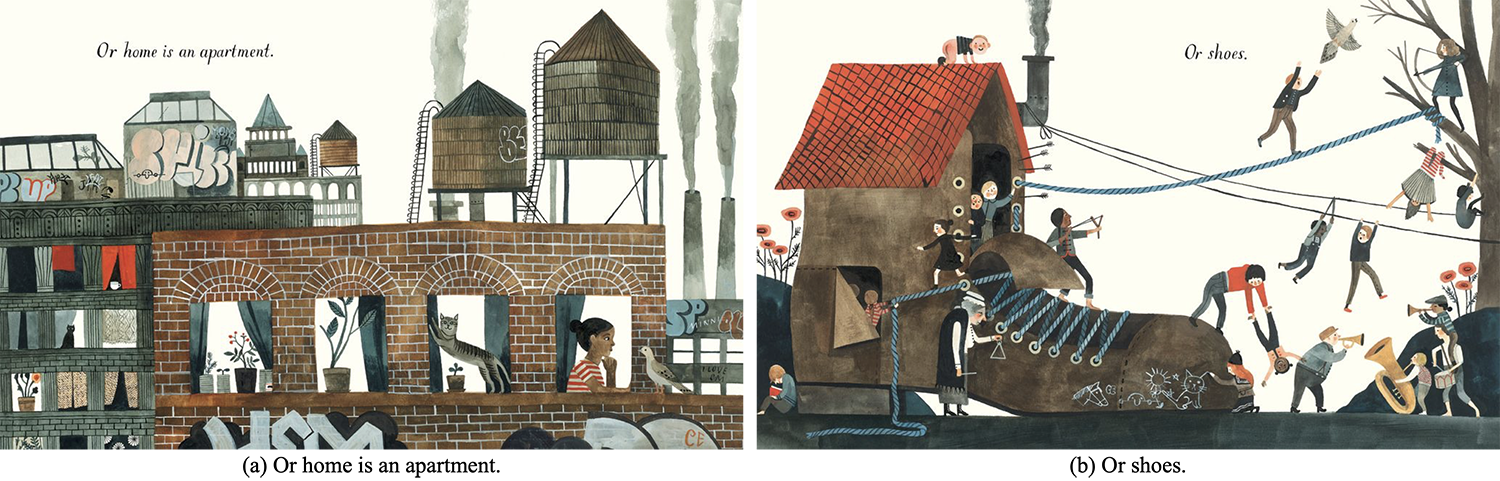}
  \caption{The intertextual relationship between complementary text and image in \textit{Home}, Credit by Carson Ellis.}
  \label{carson}
\end{figure*}

\begin{figure*}
  \centering
  \includegraphics[width=\linewidth]{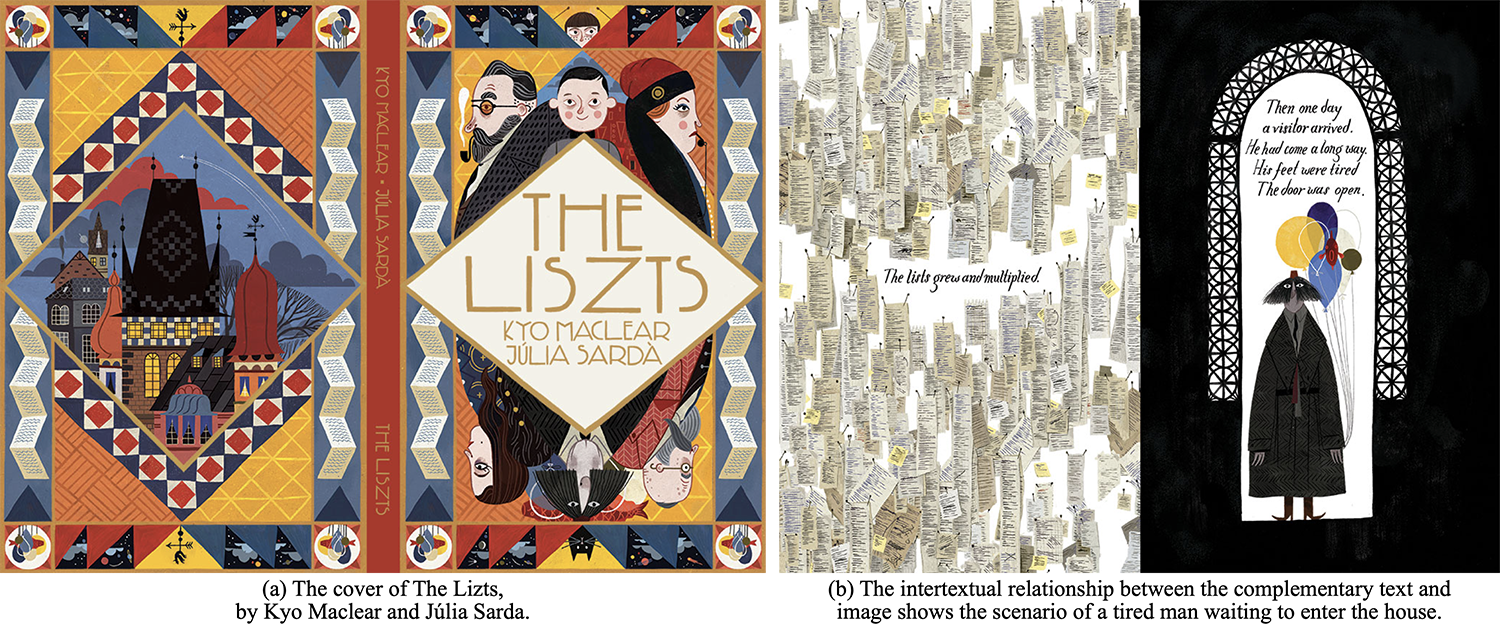}
  \caption{The intertextual relationship between complementary text and image in The Liszts, Credit by Kyo Maclear and Júlia Sarda.}
  \label{list}
\end{figure*}

While \textit{text-image intertextuality} refers to the dynamic meaning-making process that emerges from the interplay between textual and visual elements, it extends beyond simple illustration. This positions visual narratives as complex intertextual spaces where meaning arises through negotiation between what texts suggest and what images show or symbolically represent. Originating with Kristeva's work in the late 1960s, intertextuality was defined as the principle that every work is influenced by previous works~\cite{sevindik2024view,alfaro1996intertextuality}. Werner's framework identifies three categories: "within the visual," "between visuals," and "between the visual and the word" \cite{werner2004does}. We focus on the last category, encompassing "anchoring the image" (where text frames visual interpretation), "framing the word" (where visuals enhance textual credibility), and "prompting reflexivity in visuals" (where text-image combinations encourage active interpretation and meaning construction)~\cite{sevindik2024view}. However, visual narrative complexity diverges from textual logic~\cite{wolf2003narrative}. Images can simultaneously convey multiple information streams through dual logics or cyclical structures, necessitating divergent thinking in creation. Despite this, existing research typically reduces visual narrative to textual visualisation, overlooking rich intertextual relationships (see Fig.~\ref{carson} and Fig.~\ref{list}). This oversight becomes problematic for human-AI collaboration, as current tools are designed around this limited conceptualisation, lacking insights into collaborative divergent thinking stages.

\section{Related Works}


Our study addresses these gaps by empirically investigating how novice users co-create visual narratives with VLMs, focusing specifically on the emergence and negotiation of \textit{text-image intertextuality}, a dynamic, meaning-making process that remains underexplored in current human-AI collaboration research.

\subsection{Human-AI Creative Collaboration Patterns}

AI creative partnerships exhibit diverse collaboration patterns. Karimi et al. identified three creativity types (combinatorial, exploratory, transformational) in their creative sketching partner framework~\cite{karimi2020creative}, while Rezwana et al. mapped collaboration spectra from human-led to AI-led interactions~\cite{rezwana2023designing}. The degree and mode of AI integration significantly shape users' cognitive engagement~\cite{xu2025productive, progga2024large, rezwana2022understanding, fan2024contextcam, shaer2024integrating}, particularly in writing contexts where scaffolding levels affect perceived co-creative relationships~\cite{dhillon2024shaping, wan2024felt, ghajargar2022redhead, kantosalo2019quantifying}. Jakesch et al. demonstrated that users attributing greater agency to AI engage with them as creative partners rather than tools, though this creates tension with desires for creative control~\cite{jakesch2023human}. Communication modalities also affect collaboration, with verbal versus visual interactions producing distinct experiences~\cite{zhang2024verbal, zhang2023investigating, maccio2022mixed, st2015robot}. This reflects Kantosalo and Jordanous's "autonomy paradox," where users simultaneously want AI initiative while maintaining human direction~\cite{kantosalo2021role}, as echoed in writers' experiences with LLMs, where AI initiative conflicts with authorship concerns~\cite{chakrabarty2024creativity, lehmann2022suggestion, he2025contributions, marco2024pron, hutson2025human}.

In visual domains, Guzdial and Riedl's alternating contribution framework enables AI learning from human input~\cite{guzdial2019interaction}, exemplified by multimodal collaborative tools~\cite{zheng2024soap, ning2024mimosa, cohn2025multimodal}. However, turn-taking approaches may fail to adequately capture the \textit{text-image intertextuality}. For novices lacking established creative processes, AI can scaffold complex skill development~\cite{yan2025social, zhang2025breaking}, functioning as a "second mind" during ideation~\cite{wan2024felt} and empowering non-experts to manipulate AI-generated content within familiar workflows~\cite{he2023exploring}.

\subsection{VLMs for Creative Visual Narrative}

VLMs have evolved from early neural style transfer~\cite{gatys2016image} and GANs~\cite{saxena2021generative, karras2020analyzing} to current diffusion models (DALL-E3, Stable Diffusion, Midjourney) that transform noise into coherent images via text prompts, achieving unprecedented photorealism~\cite{bansal2024revolutionizing}. These advances significantly reshape both professional and amateur creative workflows~\cite{tang2024exploring}. Recent developments have further expanded capabilities through techniques such as ControlNet~\cite{zhang2023adding}, which provides more precise control over composition through sketch or pose guidance, and personalisation methods that enable systems to learn and reproduce specific concepts or styles from limited examples~\cite{sohn2023styledrop, israr2022customizing, he2024few}. Real-world implementations such as PodReels~\cite{wang2024podreels} and ReelFramer~\cite{wang2024reelframer} illustrate how collaborative human-AI systems effectively translate textual prompts into engaging multimedia narratives.

Specialised tools for visual narrative remain relatively limited despite the proliferation of general-purpose image generation systems. Some approaches address sequential narrative challenges, like Comic Diffusion~\cite{ogkalu2023comicdiffusion}, which conditions on sequential context for panel consistency, and StoryDALL-E~\cite{maharana2022storydall}, which incorporates narrative understanding into the generation process. Similarly, cross-modal generative creativity support tools, such as XCreation, facilitate advanced multimodal narrative coherence through structured graph-based interfaces~\cite{yan2023xcreation}. The interface designs also vary, ranging from conversational approaches, such as Comicchat~\cite{kurlander1996comic}, where users describe narratives in natural language, to structured interfaces like StoryDrawer~\cite{zhang2022storydrawer}, where users explicitly define characters and settings before generating sequences. Commercial applications targeting comic creation include simplified tools like Comixify~\cite{pkesko2019comixify} and more sophisticated options like Wonder Dynamics~\cite{wonderdynamics2025flowstudio}, which maintains character consistency through 3D modeling. While these systems offer increasingly sophisticated capabilities, their rapid evolution means that research findings may quickly become outdated as new techniques emerge~\cite{muller2022genaichi, amankwah2024impending, creely2025creative, ramamoorthy2025evaluating}. This underscores the importance of identifying underlying patterns and principles in human-AI collaboration that may remain relevant despite specific technological changes.

\subsection{Towards AI-Assisted Text-Image Intertextuality}

Intertextuality has long been conceptualised as the dialogic relationship between texts, where meaning emerges through the interplay of multiple voices ~\cite{drajati2023intertextuality}. With the rise of generative AI, intertextuality extends beyond verbal texts into cross-modal terrains where text and images dynamically interact. Recent work in educational writing has shown that AI can scaffold users' capacity to navigate intertextual references, though learners still struggle to establish their own voices when integrating external sources ~\cite{drajati2023intertextuality}. This suggests a broader opportunity to reconceptualize intertextuality as a multimodal process. In multimodal AI research, coherence between images and texts has been emphasised as a key to communicative effectiveness. Image–text coherence theory highlights how text may describe, explain, or evaluate imagery, while images can enrich, justify, or problematize textual meaning ~\cite{alikhani2023image}. Complementing this, inter-semiotic analyses demonstrate how text-to-image models embody ideational meanings that go beyond linear linguistic representations, raising questions about how visual and verbal semiotic systems converge or diverge in the meaning-making process ~\cite{ghazvineh2024inter}. Such findings indicate that intertextuality in AI-mediated contexts is not merely about alignment, but about constructing complementary or even contesting layers of interpretation.

Applications in education and design illustrate these dynamics. In Computer-Assisted Language Learning ~\cite{xu2024harnessing}, AI image generation offers culturally relevant, context-specific visual aids that enhance engagement with less commonly taught languages, yet also risks reinforcing stereotypes and inconsistencies ~\cite{xu2024harnessing}. In creative design workflows, text-to-image systems support early ideation by sparking imagination, though user studies indicate tensions between inspiration, control, and reflective interpretation ~\cite{liu2025exploring}. Beyond instrumental use, artistic practices employ AI-generated imagery to problematize historical memory and temporality, constructing fictive archives that both illuminate and critique cultural absences ~\cite{martin2025ai}. These examples highlight that AI-assisted intertextuality is simultaneously pragmatic, imaginative, and political. At a theoretical level, scholars have argued that intertextuality in AI art connects cognitive, psychological, and aesthetic processes, suggesting that meaning is co-constructed across disciplinary perspectives ~\cite{VWU, tilak24}. Discussions of attribution and authorship further complicate these dynamics, raising concerns over how intertextuality is acknowledged when AI reshapes semiotic resources ~\cite{rogers_2022, ni2025students}. While these debates advance our understanding, they often stop short of examining how end-users actively negotiate and reconfigure intertextuality in situated creative practice. 

Thus, prior research indicates that AI-assisted text–image intertextuality opens up new possibilities for narrative construction, education, and cultural critique, but also introduces challenges related to coherence, authorship, and user agency. Yet there remains a lack of empirical investigation into how non-expert users mobilize AI tools to expand or contest textual meanings through imagery. Our study addresses this gap by empirically analyzing human–AI co-creation to uncover how users apply visual augmentations to supplement, challenge, or reframe textual intentions.

\section{Methodology}

To explore how novice users collaborate with AI tools in visual narrative, we designed a three-phase qualitative study that combined structured task engagement, real-time observation, and reflective interviews.

\subsection{Participant Demographics and Recruitment}

Our study recruited 15 participants (8 female, 7 male, aged 23-74, M=34.9) representing diverse professional backgrounds (see Table~\ref{tab:participants}). We deliberately sought a broad participant pool that included both artistic and non-artistic backgrounds, as text–image intertextuality represents a fundamental literacy skill relevant to all individuals, not exclusively to specialised art practitioners. This pool reflected varied educational backgrounds and disciplines, but had limited experience creating visual content themselves, providing rich perspectives on visual narrative approaches. We intentionally selected participants with varying levels of experience with AI tools in image generation: 3 with no prior experience, 8 with casual usage, and 4 with moderate experience, to observe how familiarity with AI tools influenced collaboration patterns. 

All participants were native Chinese speakers, with varying proficiency in English as a second language. We provided real-time translation assistance during the experiment to maintain workflow efficiency with AI. For data analysis and reporting, we implemented back-translation procedures for all communications between the moderator and participants, as well as for all prompts, to ensure linguistic accuracy and authenticity in our findings~\cite{brislin2001back}. The study received ethical approval from our university's Institutional Review Board (IRB), and all participants provided informed consent for the recording of data and the anonymous publication of results.

\begin{table*}[htbp]
\centering
\caption{Demographic Information and Background Information of Participants. From left to right, each column presents the participant number, age, gender, profession, prior use of AI tools, and their self-reported proficiency in using AI for image generation (5 = very proficient, 1 = not proficient at all).}
\label{tab:participants}
\begin{tabularx}{0.92\textwidth}{lccccc}
\hline
\textbf{No.} & \textbf{Age} & \textbf{Gender} & \textbf{Profession} & \textbf{Experience} & \textbf{Proficiency} \\
 &  &  &  & \textbf{(AI Tools)} & \textbf{(AI Image Generation)} \\
\hline
P1 & 74 & F & Retired & None & 0 \\
P2 & 24 & M & Master Student (Design) & Intermediate & 1 \\
P3 & 24 & F & Designer & Novice & 0 \\
P4 & 50 & M & Banker & None & 0 \\
P5 & 23 & F & Master Student (Journalism) & Intermediate & 2 \\
P6 & 24 & F & Master Student (Law) & Intermediate & 2 \\
P7 & 38 & F & Officer & Intermediate & 1 \\
P8 & 24 & F & Student (Design) & Novice & 1 \\
P9 & 35 & F & Assistant Professor (Curation) & Intermediate & 1 \\
P10 & 24 & F & Student (Design) & Novice & 1 \\
P11 & 28 & M & Lecturer & Novice & 1 \\
P12 & 34 & M & Designer & Intermediate & 3 \\
P13 & 28 & M & PhD Student (Landscape Architecture) & Novice & 1 \\
P14 & 25 & M & Master Student (Design) & Intermediate & 2 \\
P15 & 36 & M & Associate Professor (Public Art) & Novice & 1 \\
\hline
\end{tabularx}
\end{table*}

\subsection{Task Design}

We designed "semi-open" and "open" tasks to observe how participants collaborated with AI image generation tools under different levels of constraint. Both task types required participants to create visual narratives comprising 6-9 related images conveying a coherent story.

\begin{enumerate}
    \item \textbf{Semi-Open Tasks:} We offered participants specific thematic constraints while allowing creative freedom in narrative development. Participants could select one task from the following options:
        \begin{itemize}
        \item \textbf{Emotional Journey of \_\_\_\_\_\_\_\_\_:} Create a visual narrative exploring a character's emotional transformation by filling in the blank with any concept, experience, or identity (e.g., "an artist," "finding purpose," "confronting fear"). \textit{The story could depict the evolution from one emotional state to another.}
        \item \textbf{Day in a Future City, I went to \_\_\_\_\_\_\_\_\_:} Visualize a sequence showing moments throughout a day in a city 100 years from now by completing the prompt with any location or activity (e.g., "the floating market," "my android therapist," "the vertical farm"). \textit{The narrative could highlight how technology has transformed everyday experiences.}
        \item \textbf{Nature's Revenge - How \_\_\_\_\_\_\_\_\_:} Tell a visual story about nature reclaiming a human-made environment by completing the prompt with a specific scenario (e.g., "vines consumed the skyscrapers," "animals transformed the mall," "the forest reclaimed the highway"). \textit{The narrative could depict the aftermath of abandonment or disaster.}
    \end{itemize}
    \item \textbf{Open Tasks:} Then, they were required to create a visual narrative on any topic of their choosing, with the only requirement being that it form a coherent sequence with a narrative arc. This task allowed us to observe how participants approached visual narrative when given complete thematic freedom.
\end{enumerate}

This dual-task approach draws from Guilford's divergent-convergent thinking framework, which describes two complementary cognitive processes used in problem-solving. Divergent thinking generates multiple unique solutions and ideas, promoting creativity and exploration. Convergent thinking, in contrast, evaluates and narrows down these ideas to arrive at a single, logical, and practical solution. Together, these thinking styles allow for both innovative idea generation and decisive problem resolution, forming an iterative cycle essential for creative problem-solving~\cite{guilford1967nature}. In visual narrative creation, this framework becomes particularly relevant as creators have to generate multiple possible storylines and visual interpretations (divergent) while simultaneously making specific compositional and narrative decisions to create coherent sequences (convergent).

We consider semi-open tasks as convergent and open tasks as divergent. Semi-open tasks provided participants with predetermined narrative elements, while open tasks offered minimal constraints. For both tasks, participants were instructed to use the AI tool to create a coherent visual narrative rather than focusing on the technical quality of individual images. We emphasised that the goal was to explore the process of visual narrative rather than to produce polished final products. This framing helped reduce performance anxiety and encouraged participants to experiment with different approaches to narrative development. This dual-task structure enables us to examine how Guilford's divergent-convergent cycle manifests differently across varying constraint levels and how AI tools may support or hinder different phases of this creative thinking process in visual narrative development. As Davis has observed, creative constraints can sometimes facilitate rather than hinder creative expression, particularly for novices~\cite{davis2018flexibility}. By including both task types, we could observe whether participants' collaboration strategies and challenges differed between more and less constrained contexts.

\subsection{Materials and Tools}

Our study employed GPT-4o\footnote{At the time of data collection (March 2025), we utilised GPT-4o as a tool for this study. While more advanced models have since emerged (including GPT-5 and specialised image generation tools like Nano Banana), our focus on \textit{text-image intertextuality} and collaborative narrative development processes remains relevant across different AI platforms. The specific technical capabilities of image generation were less critical to our research objectives than understanding how creators negotiate meaning-making between textual prompts and visual outputs in collaborative narrative contexts.} as the primary collaborative tool. This integrated VLM offered a unified interface for text and image creation, allowing participants to describe narrative concepts through natural language prompts and iteratively refine generated images through conversation. Participants articulated complete narrative concepts guided by the moderator before image generation to maintain coherent visual narrative logic and ensure continuity across iterations. All participants accessed standardised accounts under controlled conditions to ensure methodological consistency.

\subsection{Research Procedure}

\begin{figure*}
  \centering
  \includegraphics[width=\linewidth]{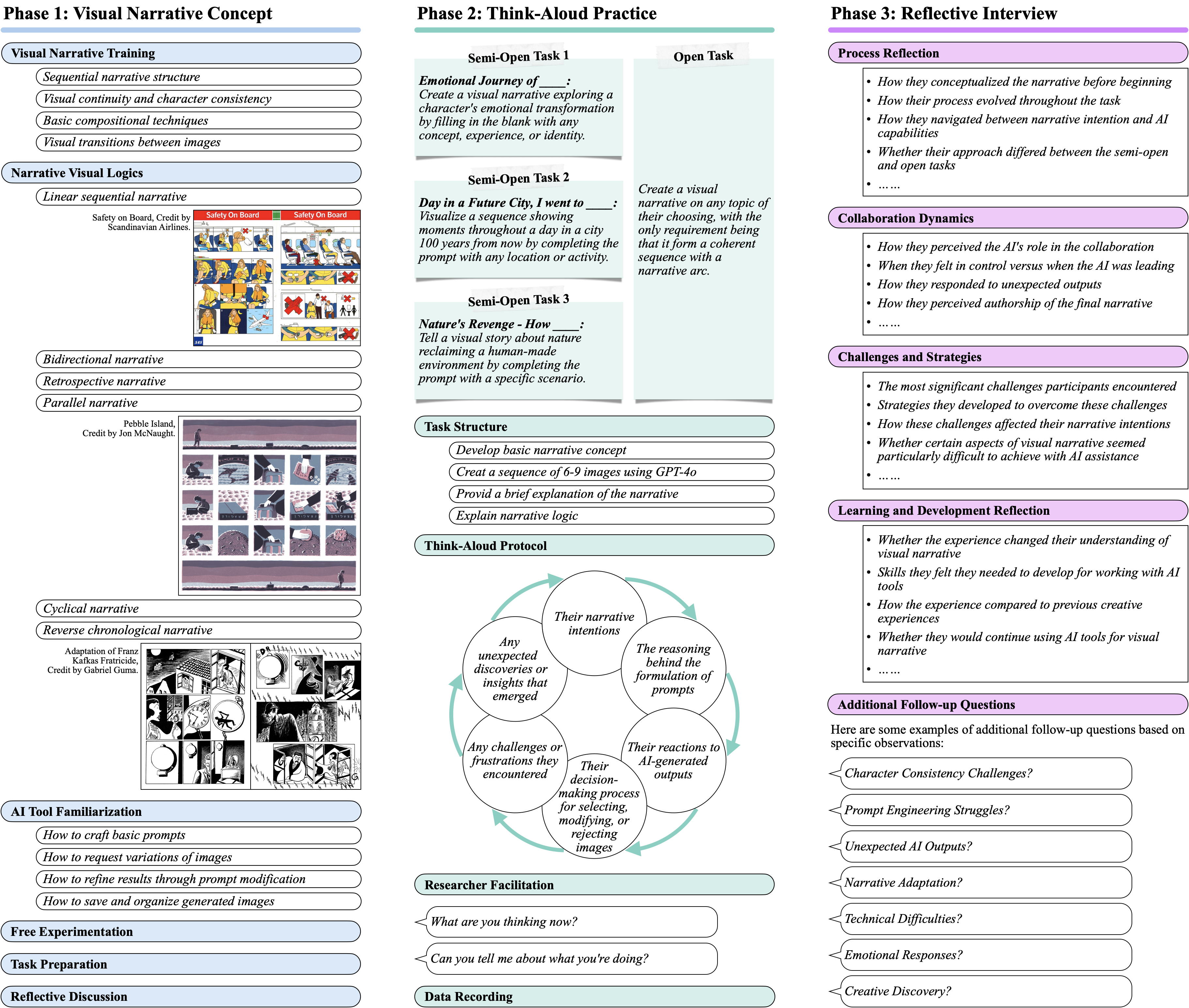}
  \caption{Our research phases, Credit by Authors.}
  \label{studyflow}
\end{figure*}

Our research procedure consisted of three sequential phases designed to capture the \textit{text-image intertextuality} of the human-AI collaborative process in visual narratives (see Fig.~\ref{studyflow}).

\subsubsection{Phase 1: Visual Narrative Concept}

\begin{figure*}
  \centering
  \includegraphics[width=\linewidth]{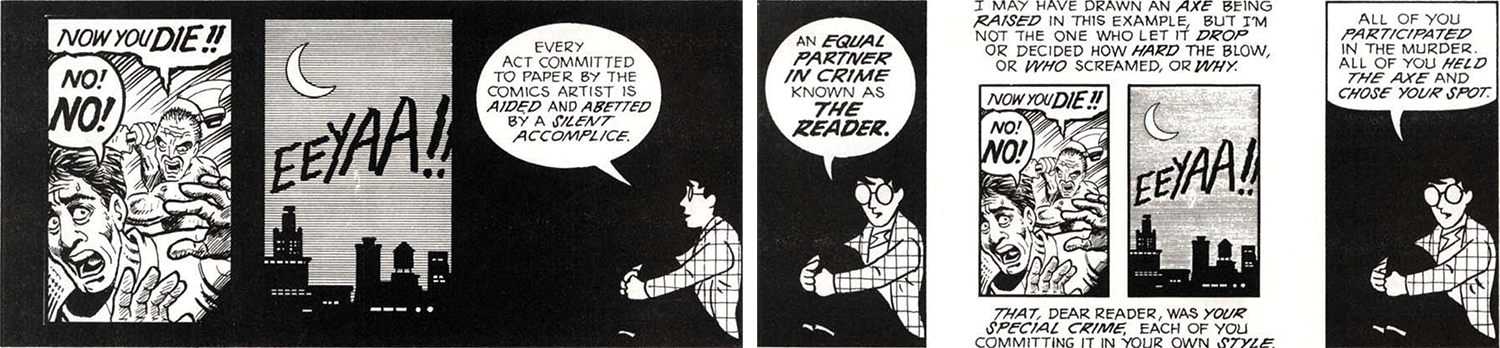}
  \caption{Users automatically complete their given storylines in sequential images in Understanding Comics~\cite{mccloud1993understanding}, Credit by Scott McCloud.}
  \label{bw}
\end{figure*}

\begin{figure*}
  \centering
  \includegraphics[width=\linewidth]{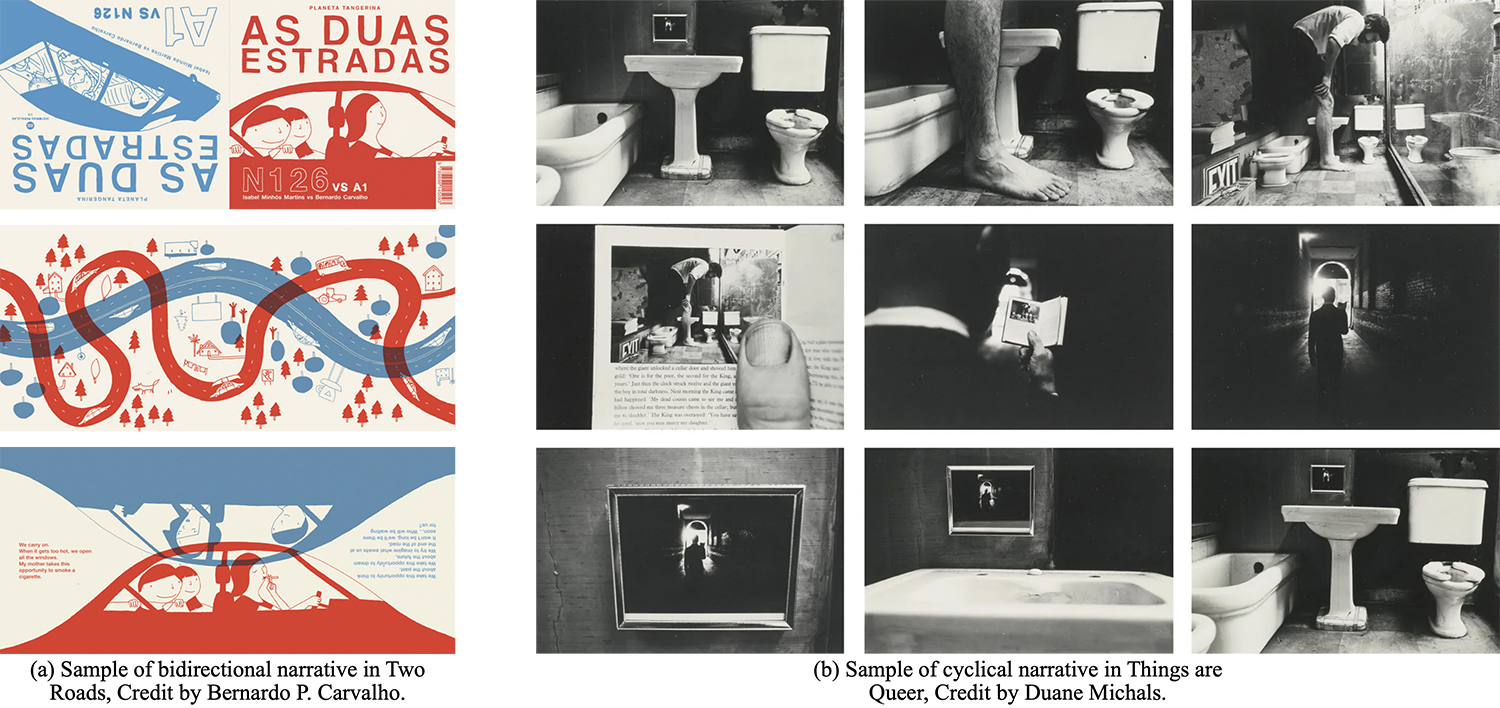}
  \caption{Sample of bidirectional narrative and cyclical narrative, Credit by artists.}
  \label{two}
\end{figure*}

The first phase introduced participants to basic concepts of visual narrative and familiarised them with the AI image generation tools they would be using. First, we provided a brief introduction to key visual narrative principles, including sequential narrative structure (beginning, middle, end), visual continuity and character consistency, basic compositional techniques (framing, perspective, and focus), and visual transitions between images (see Fig.~\ref{bw}). Then we introduced several narrative visual logics, including linear sequential narrative, bidirectional narrative (see Fig.~\ref{two}a), retrospective narrative, parallel narrative, cyclical narrative (see Fig.~\ref{two}b), and reverse chronological narrative, etc. Second, participants were introduced to the text-to-image generation tools they would be using (GPT-4o accessed through web interfaces). We demonstrated basic functionality, including how to craft basic prompts, how to request variations of images, how to refine results through prompt modification, and how to save and organize generated images. They had 10-15 minutes to experiment freely with the tools, generating sample images based on simple prompts of their choosing. Finally, we encouraged them to sketch brief narrative outlines or key points they wanted to include in their visual story and consider choosing one semi-open task. 

During this process, we asked participants to discuss their thoughts with us, providing insight into their initial conceptualisation of the task and their expectations for the AI collaboration. This phase laid the foundation for subsequent tasks, providing us with baseline observations of participants' understanding of visual narrative and their initial approach to working with the AI tool.

\subsubsection{Phase 2: Think-Aloud Practice}

The second phase constituted the core of our study, with participants completing the given two tasks while thinking aloud throughout the process~\cite{tomitsch2020}. This phase employed a standard think-aloud protocol~\cite{ericsson1998study}, with participants expressing their thoughts, decisions, and reactions as they worked. According to Tomitsch~\cite{tomitsch2020}:

\begin{quote}
    \textit{"it encourages people to verbalise what they are thinking as they perform a task, revealing their cognitive processes (p. 124)."} 
\end{quote}

For both tasks we designed, participants developed their basic narrative concept, created a sequence of 6-9 images using GPT-4o for each task, provided a brief explanation of their narrative, and explained their narrative logic to us. Throughout this process, participants were encouraged to verbalize their narrative intentions, the reasoning behind the formulation of prompts, their reactions to AI-generated outputs, their decision-making process for selecting, modifying, or rejecting images, any challenges or frustrations they encountered, and any unexpected discoveries or insights that emerged. When participants fell silent for extended periods, researchers provided minimal prompts such as "\textit{What are you thinking now?}" or "\textit{Can you tell me about what you're doing?}" to encourage continued verbalisation. All screen activity was recorded using screen capture software, providing a complete record of the interaction with GPT-4o, including all prompts entered, images generated, and the selection/rejection process. This visual record complemented the audio recording of participants' think-aloud narration, allowing us to conduct a comprehensive analysis of the collaborative process.

\subsubsection{Phase 3: Reflective Interview}

The final session consisted of a semi-structured interview prompting participants to reflect on their GPT-4o collaboration experience across four areas: 1) process reflection (overall approach, narrative conceptualisation, evolution, and navigation between intention and AI capabilities), 2) collaboration dynamics (AI's perceived role, control dynamics, responses to unexpected outputs, and authorship perceptions), 3) challenges and strategies (significant obstacles, developed solutions, and effects on narrative intentions), and 4) learning and development reflection (changed understanding, needed skills, and future AI usage). The protocol included core questions for all participants, as well as targeted follow-up questions based on specific observations from the think-aloud phase, providing insights into participants' subjective collaboration experiences that complemented the observational data.

\subsection{Data Collection}

We collected data through several distinct phases using complementary instruments to capture the full spectrum of participant experiences. Prior to the study, we administered a questionnaire to gather participant demographics, creative backgrounds, and prior experience with AI image generation tools, establishing baseline characteristics for each participant.

During Phase 1, we collected data from informal discussions designed to build rapport, help participants become comfortable with the technology, and provide initial insights into participants' creative approaches and expectations. In Phase 2, we documented key prompt strategies, participant reactions to AI outputs, collaboration challenges, and creative breakthroughs through real-time observation notes. We also preserved complete conversation logs containing all prompts, AI responses, and generated images with precise timestamps to capture the chronological development of each narrative. Phase 3 data collection involved post-creation semi-structured interviews following a protocol that included both standard questions for all participants and session-specific follow-up questions based on observations from Phase 2.

Following all phases, we compiled the multi-source data, including questionnaire responses, observational notes, conversation logs, generated images, interview transcripts, and final assessments. This comprehensive data collection approach enabled methodological triangulation, allowing us to use multiple data sources and collection methods to validate and enhance the credibility of our findings. By cross-referencing participants' stated intentions from interviews, observed behaviors during creation sessions, actual AI interactions from conversation logs, and reflective insights from questionnaires, we developed a robust understanding of the human-AI creative collaboration process while identifying both consistencies and discrepancies across different data sources.

\section{Data Analysis}

\subsection{Multi-Perspective Analysis}

Our analytical approach followed a comprehensive research process of theoretical guidance → inductive discovery → structured validation, where Creative Collaboration Theory provided theoretical sensitivity and conceptual foundation, Constructivist Grounded Theory served as the primary method ensuring data-driven findings, and fsQCA functioned as supplementary analysis to validate and deepen our grounded theory discoveries.

Based on Creative Collaboration Theory~\cite{sawyer2017group, John-Steiner2000}, we approached our analysis with an understanding of how collaborative creative processes unfold through negotiated roles, shared understanding, and emergent creative outcomes. This theoretical framework provided essential conceptual sensitivity for examining human-AI partnerships, particularly in identifying moments of role negotiation, integrating creative input, and collaborative problem-solving during the co-creation of visual narratives. The Creative Collaboration Theory guided our initial approach to understanding the unique dynamics where humans and AI systems contribute different forms of creative input and develop a shared understanding through iterative exchanges.

We applied Constructivist Grounded Theory~\cite{charmaz2006constructing} as our primary analytical methodology for systematically analyzing the rich multimodal data collected from our three-phase qualitative study. This approach was particularly well-suited for several reasons. First, Constructivist Grounded Theory excels at exploring processes and interactions in emerging domains where existing theoretical frameworks may be insufficient. As Charmaz notes~\cite{charmaz2012qualitative}, this methodology \textit{"places priority on the studied phenomenon and sees both data and analysis as created from shared experiences and relationships with participants" (p. 349)}, a perspective that acknowledges the co-constructed nature of understanding in novel contexts like human-AI co-creation. Second, this approach enabled systematic analysis while remaining sensitive to participants' subjective experiences and the sociocultural contexts that shaped them. Given the diversity of our participant pool, this sensitivity to individual perspectives and contexts was essential for developing detailed insights into the collaborative process. Third, the iterative approach enabled us to refine our focus and follow emerging themes as the study progressed, allowing for theory development that was truly grounded in participant experiences.

We then used fuzzy-set Qualitative Comparative Analysis (fsQCA)~\cite{ragin2009redesigning,zschoch2011configurational} as supplementary analysis to systematically examine the complex combinations of conditions that led to different collaboration outcomes. After completing our grounded theory analysis, we identified six key variables that influence collaboration outcomes: collaborative mindset, script-first approach, iteration willingness, technical knowledge, visual literacy, and educational perspective. We then calibrated these conditions into fuzzy-set membership scores (0.0-1.0) for each case and conducted necessity and sufficiency analyses to identify pathways to successful intertextual narrative creation. This method provided a structured approach to validate and extend our grounded theory findings by moving beyond identifying individual factors to understanding how these factors interact to produce successful visual narratives.

\subsection{Coding Process and Results}

Our coding process followed a methodological framework guided by Creative Collaboration Theory through inductive discovery to structured validation. We analysed data from 15 participants, including 22.5 hours of transcribed recordings, 48 visual narratives (each agreed story was assigned as one piece), 105 rejected images, 89 pages of interview notes, and 224 translated script materials.

Following Charmaz's constructivist grounded theory approach~\cite{charmaz2006constructing}, two researchers conducted initial coding using participants' backgrounds and expectations to develop sensitizing concepts. Two additional researchers independently performed line-by-line analysis of interviews, marking video content in 2-3 minute segments and using open coding to capture specific human-AI interaction behaviors. Codes were formulated as verbs (e.g., "adjusting prompt strategies," "questioning AI outputs," "integrating fragmented ideas") rather than descriptive labels. We employed the constant comparative method, continuously comparing new data with existing codes until consensus was reached.

\begin{figure*}
  \centering
  \includegraphics[width=\linewidth]{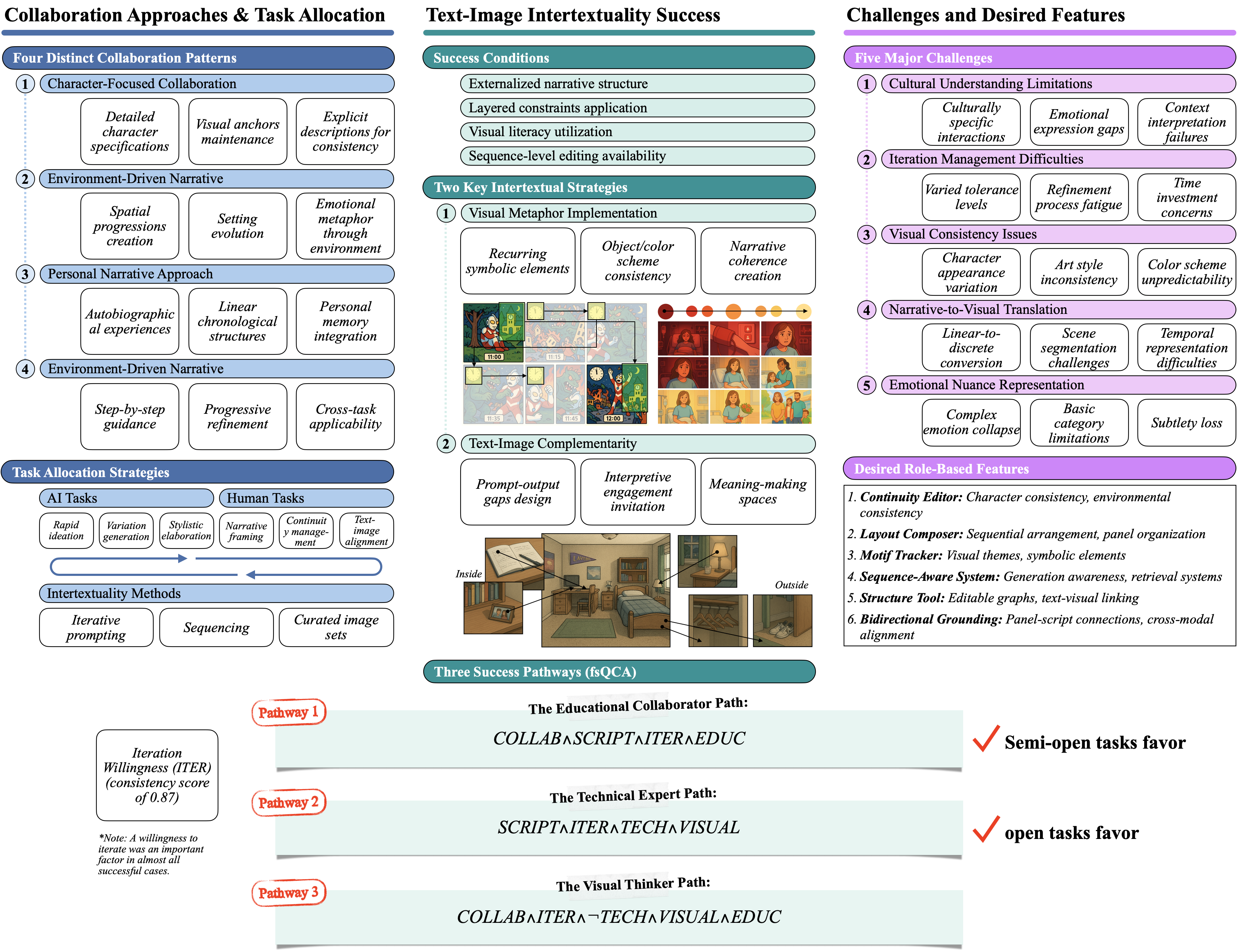}
  \caption{Overview of our study's findings. We summarised three themes aligned with the research questions: collaboration approaches \& task allocation, text-image intertextuality success, challenges and desired features, including seven sub-themes and key success pathways.}
  \label{finding}
\end{figure*}

For \textbf{RQ1}, we identified four distinct collaboration patterns documenting how AI handles rapid ideation while humans maintain narrative control. For \textbf{RQ2}, we established success conditions and documented two key intertextual strategies, identifying three pathways to successful collaboration through fsQCA. For \textbf{RQ3}, we identified five major challenges and six desired role-based AI features to support intertextuality workflows (see Fig.~\ref{finding}).

\section{Finding 1: Collaboration Approaches and Task Allocation (RQ1)}

In the study, we observed various user-directed collaboration strategies as participants engaged with AI across two tasks. Our analysis identified four distinct patterns of collaboration.

\subsection{Four Distinct Collaboration Patterns}

Our analysis identified four distinct patterns of collaboration, and the following sections provide a detailed explanation of these patterns.

\subsubsection{\textbf{Pattern 1: Character-Focused Collaboration}}

When participants (P2, P8, P11, P12, P13) engaged with the semi-open task, they typically filled in gaps directly to develop their storylines or solicited concise ideas from AI rather than offering longer prompts. Examples include P2's straightforward "Emotional Journey of Dog," which contrasts with P4's more poetic "The Emotional Journey with the Most Important Person in My Life." These simpler, direct vocabulary inputs enabled the AI to grasp storylines more intuitively, resulting in clearer guidance for specific image content at each step compared to P4's more complex narrative approach. Comparable outcomes have been observed in child-AI co-creative narrative contexts, where direct, simple prompts from children facilitated effective narrative construction~\cite{zhang2024mathemyths}.

\begin{figure*}
  \centering
  \includegraphics[width=\linewidth]{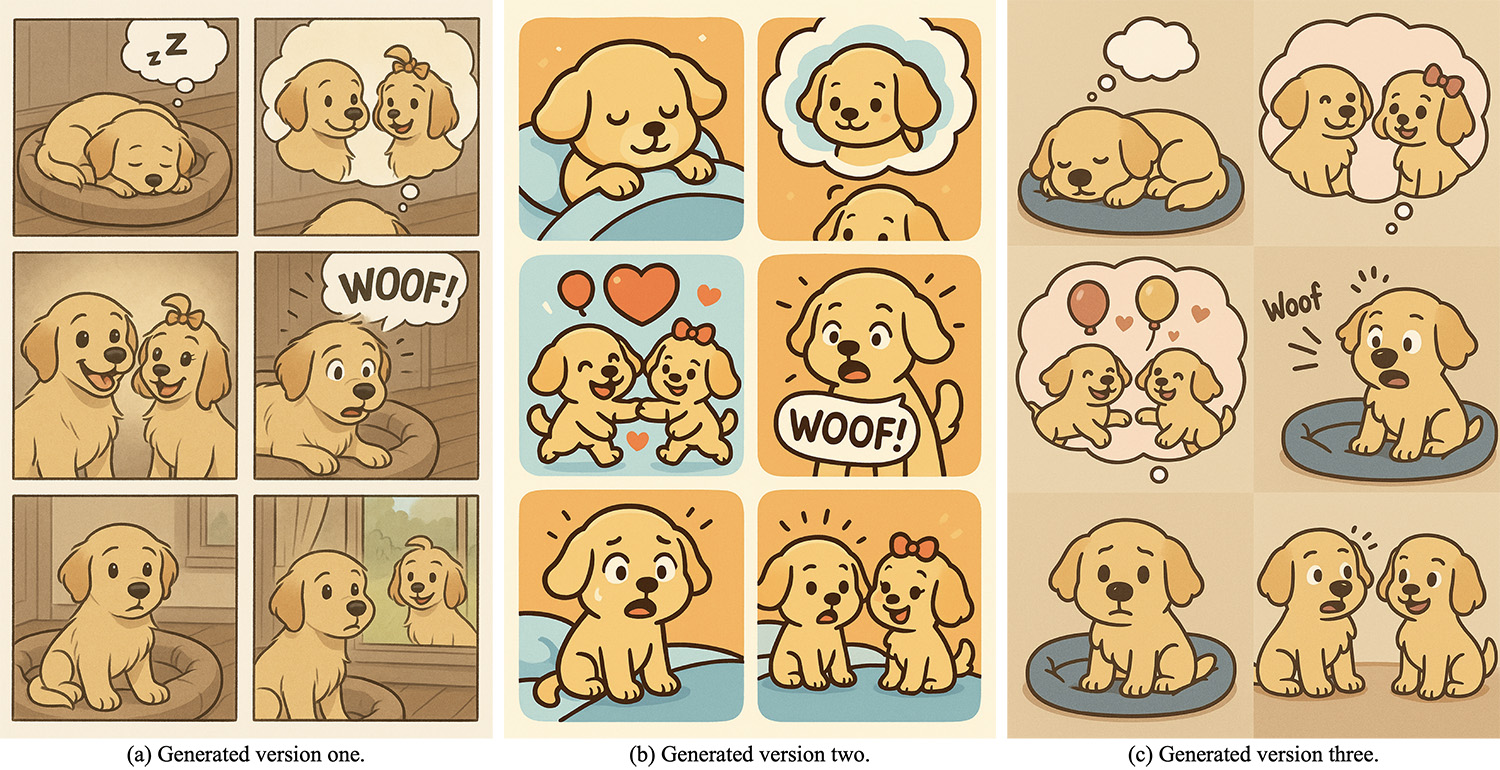}
  \caption{The three stories generated by P2. In Picture 4 of all three versions, the AI depicted the scene as the main character barking instead of hearing another dog's bark. Meanwhile, the bow on the female dog sometimes appeared and sometimes disappeared.}
  \label{p2}
\end{figure*}

For example, when tackling the "Emotional Journey of \_\_\_\_\_\_\_\_\_" semi-open task during Think-Aloud Practice, P2 took highly specific character descriptions as their central strategy: \textit{"I want to write about a dog's emotional journey, because I think this would be an interesting topic."} From the outset, he had a clear storyline in mind: a golden retriever dreaming about a female golden retriever, being awakened, and then discovering the dream had become reality. This narrative framework was fully formed before engaging with the AI. By providing detailed specifications such as: \textit{"I think I need to add a sentence to the first image, I need to add that the overall style should be cute and cartoon-like."} And he required consistent appearance across frames with more specific emotional and visual elements, he established strong visual anchors throughout their narratives (see Fig.~\ref{p2}): \textit{"In the third image, it's chasing balloons with the female golden retriever in the dream, bubbling with happiness, emitting sparks of hearts, the two dogs are very happy."}

The initial prompt he offered to AI was as follows: 

\begin{quote}
\textit{"Help me generate a storyboard with six storyboards. Picture 1: A little golden retriever had a dream; Picture 2: The little golden retriever dreamed of a cute little golden retriever girl in his dream; Picture 3: He was very happy with the little golden retriever girl in his dream; Picture 4: Suddenly he heard a dog barking, waking up from his sweet dream; Picture 5: He opened his eyes and looked forward, wondering what woke him up; Picture 6: He found out that it was the little golden retriever girl in his dream."}
\end{quote}

During the Reflective Interview, P2 emphasised the importance of pre-conceptualizing characters: \textit{"Having a clear image in mind first helps AI maintain consistency."} This approach proved effective for maintaining character consistency but sometimes limited emotional expressiveness. He also showed consideration of the sequential understanding limitations: the AI seemed to treat each frame as somewhat independent, rather than understanding the overall narrative context. This was evident in the sound effect visualisation misinterpretation in all generated 3 stories: \textit{"In the fourth picture, suddenly, it heard a dog bark, which woke it up from its beautiful dream."} The AI rendered this as the main character barking rather than hearing another dog's bark.

Another interesting phenomenon emerged in P2's second story (see Fig.~\ref{p2}b), where the AI implemented distinct color schemes to differentiate between reality and dream states, a visual technique that he had not explicitly requested in their prompts. During artifact analysis, we observed how the AI-generated images used warmer orange tones for scenes set in reality and cooler blue hues for the dream sequence where the protagonist dog encounters its canine love interest. This color differentiation effectively communicated the transition between consciousness states to viewers without requiring explicit textual explanation. While we cannot definitively determine whether this color distinction was an intentional narrative device applied by the AI or simply an emergent pattern arising from its training data, the effect nonetheless successfully enhanced the narrative. P2 confirmed during Phase 3 that he hadn't consciously considered using color differentiation in the prompt: \textit{"I didn't specifically ask for different colors for different states of consciousness."} 

This case illustrates how AI might draw from visual narrative conventions in its training data to produce visually coherent narratives, sometimes introducing elements that unexpectedly complement the human collaborator's intentions.

\subsubsection{\textbf{Pattern 2: Environment-Driven narrative}}

\begin{figure*}
  \centering
  \includegraphics[width=\linewidth]{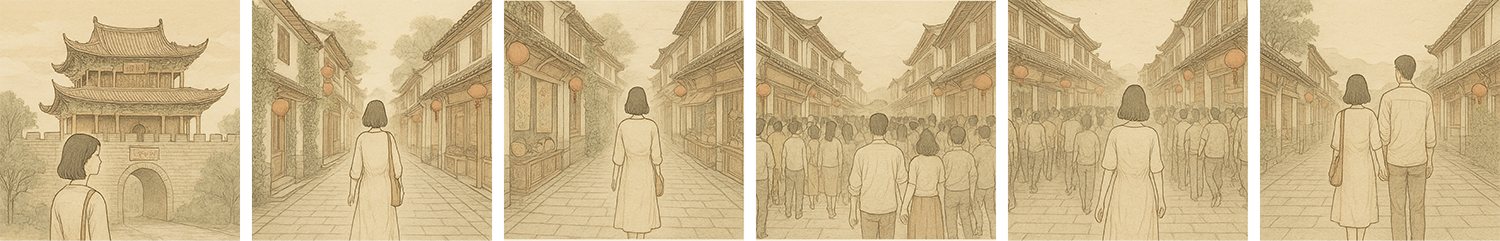}
  \caption{The generated story by P4. However, characters still appeared in the images, and they occupied the central position. There is a meaningful interphotographic relationship between the second and the fifth images. The former only shows the protagonist and the background, while the latter adds many pedestrians on this basis, which can be regarded as a manifestation of loneliness.}
  \label{p4}
\end{figure*}

For the "Day in a Future City, I went to \_\_\_\_\_\_\_\_\_" and "Nature's Revenge - How \_\_\_\_\_\_\_\_\_" semi-open tasks, environment-focused participants (P1, P3, P4, P5, P9, P10, P15) created spatial progressions to convey narrative development, while the rest of participants centred on events or activities, such as "Day in a Future City, I went to Three Thousand Years Ago" by P6, "Nature's Revenge - How Animals Defeated Machines" by P7, and "Day in the future city, I dig a hole in the universe" by P14. Unlike participants who prioritised applying one-word answers to fill in the gaps, such as P3's "A Day in the Future City, I went to Underground", P5's "A Day in the Future City, I went to Mars", as well as P10's "Nature's Revenge - How to Destroy a City", P4 articulated an environmental progression approach to visual narrative during his prompt (see Fig.~\ref{p4}) in semi-open task "The Emotional Journey with the Most Important Person in My Life": 

\begin{quote}
\textit{"I'm going to write a story titled 'Emotional Journey with the Most Important Person in My Life.' I'd like you to help me express the beautiful memories during this journey. The character is a woman, the location is Dali Ancient City, experiencing the warmth of life. My idea is: the first image shows the ancient city, the second shows the street, the third shows shops, the fourth shows crowds, the fifth shows holding hands, and the sixth shows moving forward. I need you to refine all the scenes I've described and create the images."}
\end{quote}

His approach prioritised spatial and environmental evolution over character development, using setting progression as a metaphor for the development of relationships. During the reflective interview, he explained this decision: \textit{"For example, when I said 'ancient city,' I just wanted the ancient city without figures, so it might not include any people, just an ancient city. Then, for the street, I wanted an ancient city street, also without figures, so it might just show an ancient city street..."} This environment-centered approach allowed participants to use setting evolution as an emotional metaphor, though AI struggled with the abstract-to-concrete visual translation, as he noted: \textit{"The gap is huge; for ancient city, it's just a city tower... but what I wanted to express was an overall aerial view of the entire ancient city."} 

This highlights the limitations of human-AI visual co-creation, where significant conceptual gaps persist between human intentions and AI interpretations. Future development of co-creative AI systems could address these conceptual-visual translation gaps to better support environment-driven narrative approaches. However, if we analyze these generated pictures from a formal aesthetic perspective, we can imbue them with a poetic connotation: She (the character in the sequence) stepped into the desolate ancient city and lingered there for ages. As the throngs grew, he (the male figure in picture 4) reached for her hand, and they strolled amidst the crowd. In the end, it was as though the crowd melted away, leaving only the two of them in their secluded world. This reflects what we believe to be the human use of creativity in interpreting AI-generated images, showcasing the intertextuality and values between text and image.

\subsubsection{\textbf{Pattern 3: Personal Narrative Approach}}

\begin{figure*}
  \centering
  \includegraphics[width=\linewidth]{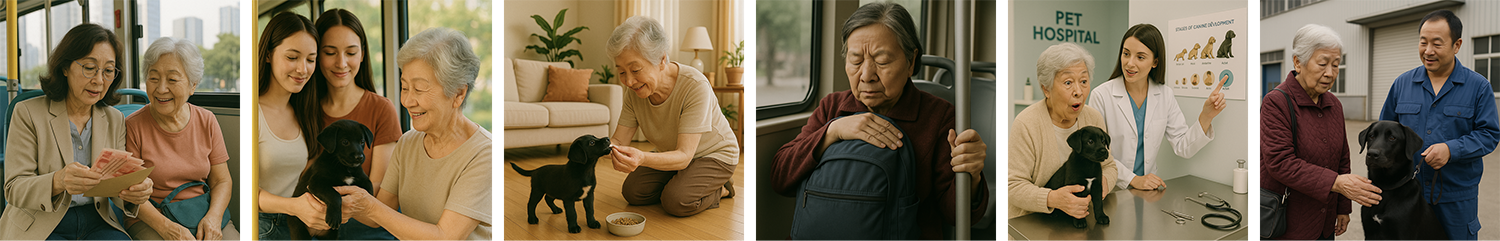}
  \caption{The story generated by P1, which is based on a real-life narrative set in the 1970s. While the story is coherent, the characters lack consistency.}
  \label{p1}
\end{figure*}

When given complete creative freedom in the open task, four participants (P1, P5, P9, P11) drew inspiration from personal experiences rather than imagination. P1 chose to tell the story of finding a stray dog (see Fig.~\ref{p1}): \textit{"I want to tell the story of how I took in a stray dog. One day, I went out with my classmate to Gu'an (a county in China, Hebei Province). I found a stray dog."} Such autobiographical approaches typically followed linear chronological structures, which participants found natural for narrative but challenging to translate into visual sequences. For example, P1 used "Square 1:1" as her leading content with each prompt to recall her memorable story of the stray dog, as follows:

\begin{quote}
\textit{"Square 1:1, Two girls in their 70s took a bus to Gu'an. Their main task was to pay the rent. They took a bus for more than an hour.}

\textit{Square 1:1, Grandma put the dog in her schoolbag to stop the puppy from barking. Pets were not allowed on the bus, but the puppy was less than a month old and very small. Grandma took the risk and put the puppy in her schoolbag and brought the puppy back to Beijing and back home.}

\textit{Square 1:1, The puppy raised by my grandma is very smart and beautiful. My family likes it very much. When feeding it, no matter where you put it, it can find the food.}

\textit{Square 1:1, Grandma takes the dog on the bus, grandma hides the dog in her schoolbag and holds it in her arms, not letting it bark, not letting the dog show up, not letting others see the dog.}

\textit{Square 1:1, Grandma took the puppy to the pet hospital for a physical examination, and the result was very good, but the doctor said that the little black dog will grow into a large dog, and it is not suitable for the elderly to adopt it."}
\end{quote}

\begin{figure*}
  \centering
  \includegraphics[width=0.25\linewidth]{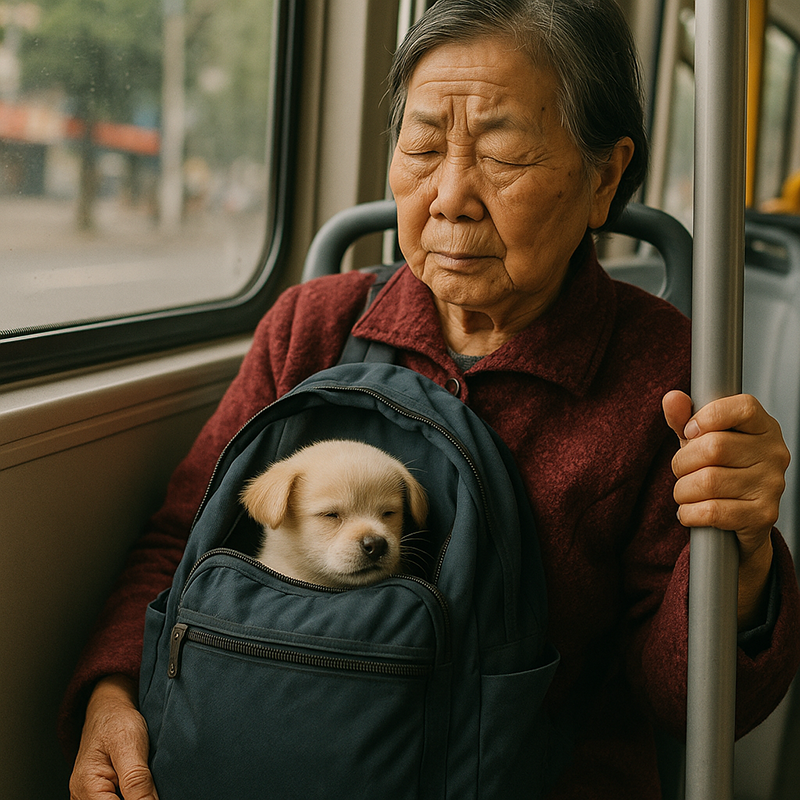}
  \caption{The generated image shows a white dog instead of a black one by P1.}
  \label{p1no}
\end{figure*}

However, P1 demonstrated limited adaptability to AI when using personal narratives as the foundation for her story. When the AI-generated images didn't match her intentions (such as showing a white dog instead of a black one, or showing the dog's head exposed from the backpack) (see Fig.~\ref{p1no}), she expressed dissatisfaction but struggled to reformulate her prompts to correct these issues. This highlights a critical challenge: when participants base their stories on personal memories, they reference specific, non-fictional visual expectations derived from their lived experiences. The resulting discrepancies between remembered images and AI-generated representations often triggered initial disappointment. Nevertheless, P1 ultimately expressed overall satisfaction with the results, suggesting a cognitive adjustment process wherein participants recognised that AI functions not as a memory repository but as an interpretive tool that can recreate situations rather than reproduce actual memories. This acceptance behavior demonstrates how participants negotiate the tension between personal visual memories and AI's interpretive capabilities, ultimately adapting their expectations to accommodate the generative rather than reproductive nature of AI-assisted visual narratives.

\subsubsection{\textbf{Pattern 4: Iterative Refinement in Complex Visual Narratives}}

\begin{figure*}
  \centering
  \includegraphics[width=0.7\linewidth]{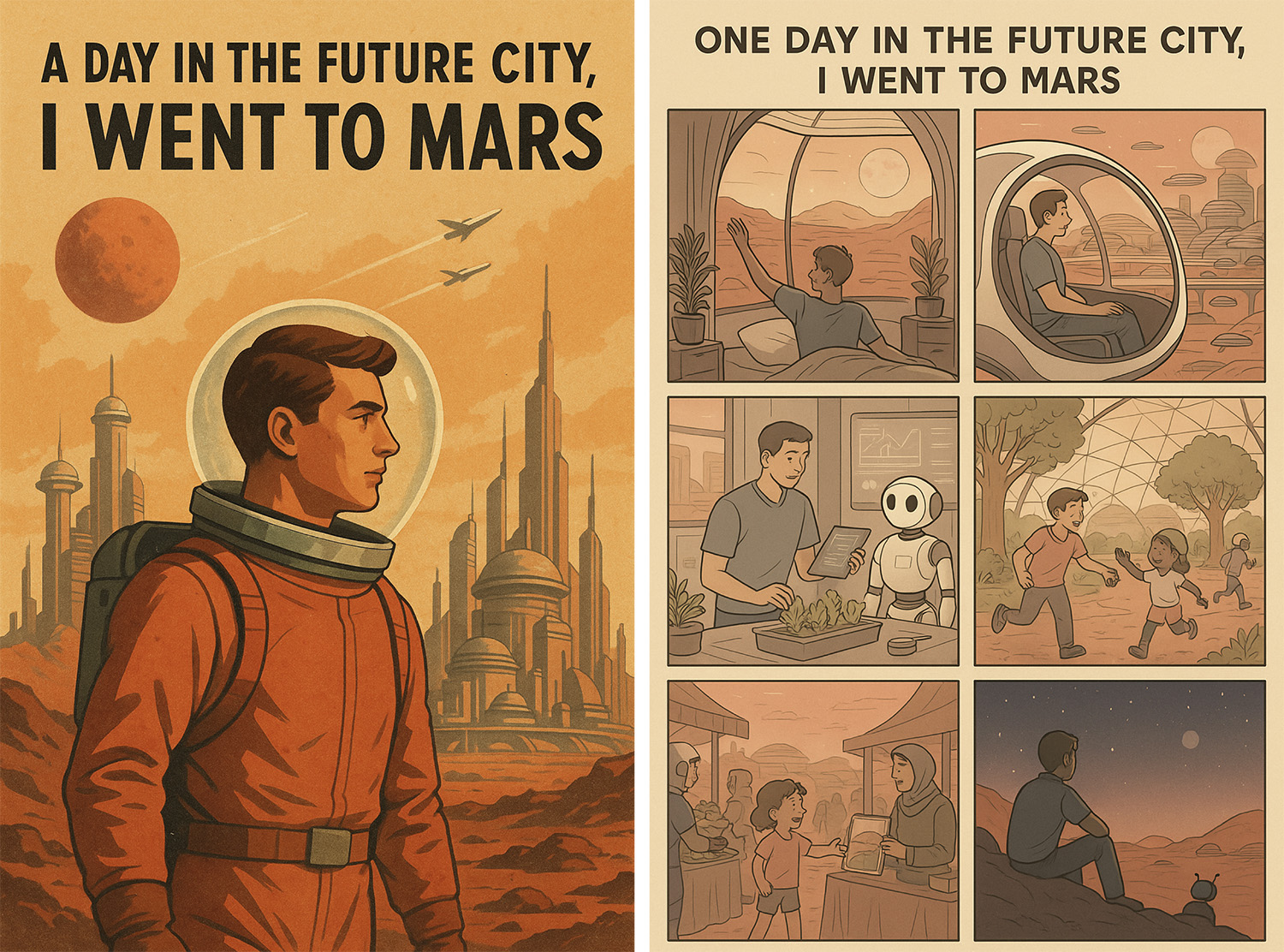}
  \caption{The generated story of "One day in the future city, I went to Mars" by P5.}
  \label{p5}
\end{figure*}

Participants (P4, P5, P6, P8, P12, P14) adopting process-oriented approaches demonstrated the most systematic workflows across both task types. These participants emphasised step-by-step guidance and progressive refinement. Throughout the tasks, P5 emphasised the importance of gradually guiding the AI rather than expecting perfect results from initial prompts (see Fig.~\ref{p5}): \textit{"The initial script and instructions were a bit too rushed, rather than guiding step by step.}" At the end of tasks, it became a central insight for her by the end of the session: "\textit{After using it this time, what's most profound for me is that I still need to guide it step by step."} The prompt she offered for the semi-open task:

\begin{quote}
\textit{"Please help me generate a 6-storyboard for a story. The title of the story is 'One day in the future city, I went to Mars.' I hope to draw the appearance of the city on Mars 100 years from now, and the day when humans live there 100 years from now, referring to the current urban construction. Please help me think about what activities will be included in this day and how to depict them through 6 storyboards."}
\end{quote}

\begin{figure*}
  \centering
  \includegraphics[width=\linewidth]{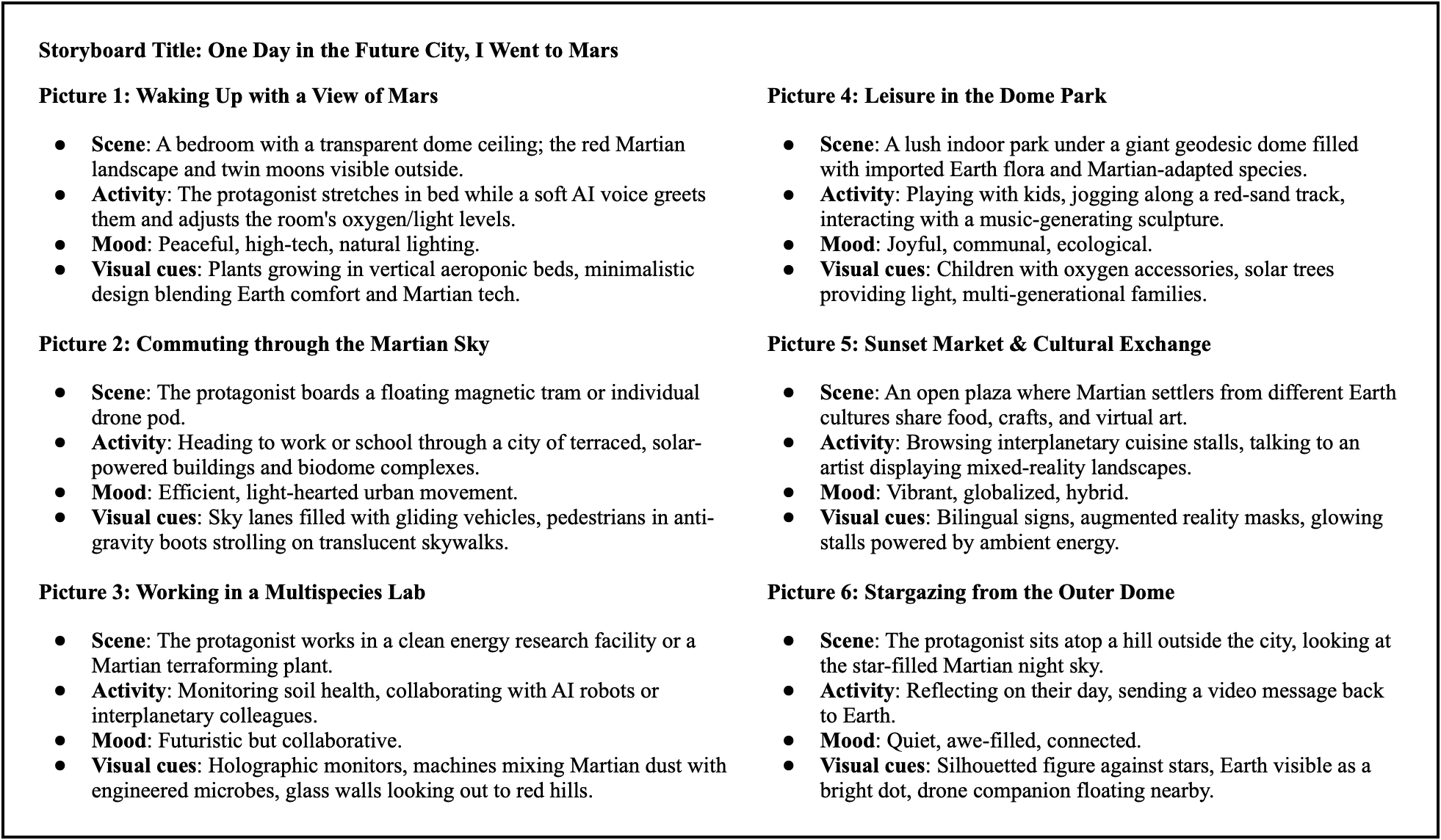}
  \caption{The AI response to P5.}
  \label{p51}
\end{figure*}

The AI responded to her with the following content, providing detailed information for each image (see Fig.~\ref{p51}). These participants showed greater patience with AI feedback, viewing iterations as a learning opportunity. As P5 developed a sophisticated view of AI as a tool that both requires skill to use effectively and helps develop that skill through interaction: \textit{"So now I see AI as a process of exercising myself; using it can actually train my expression very well. Or some descriptions of things. My feeling is that this is also a new learning process."}

\subsection{Task Allocation Strategies}

We also find task allocation strategies where participants naturally developed complementary roles, leveraging each agent's comparative advantages. Participants utilised VLMs for rapid ideation to generate diverse concepts quickly, variation generation to produce multiple visual interpretations, and stylistic elaboration to develop atmospheric details and artistic refinements. Humans maintained control over narrative framing to establish story structure, continuity management to ensure consistency across panels, and text-image alignment to create meaningful relationships beyond literal illustration. Users pursued intertextuality through iterative prompting, which systematically refined AI instructions based on evaluations of the output. This process involved sequencing strategies that established narrative flow across multiple panels and curated image sets, where participants generated multiple options and strategically selected combinations that enhanced narrative meaning through visual juxtaposition and thematic resonance.

\section{Finding 2: Text-Image Intertextuality Success (RQ2)}

Our analysis also illustrated distinct strategies participants employed to establish intertextual connections between visuals and narrative elements when co-creating with AI. 

\subsection{Success Conditions}

Participants' success in human-AI visual narrative collaboration occurred when they externalised narrative structure and layered constraints, creating explicit frameworks that guided AI generation while maintaining creative control. Effectiveness varied significantly with participants' visual literacy levels and their access to sequence-level editing capabilities, suggesting that both conceptual understanding and technical affordances influenced collaborative outcomes. The useful intertextuality emerged through selective curation and iterative revision processes, where participants generated multiple visual options and systematically refined both prompts and selections. This approach created meaningful text-image relationships that transcended simple illustration, ultimately producing coherent visual narratives through strategic human oversight combined with AI computational capabilities.

\subsection{Two Key Intertextual Strategies}

We identified two noteworthy strategies that demonstrate how participants attempted to bridge the conceptual and visual domains.

\subsubsection{\textbf{Strategy 1: Visual Metaphor Implementation}}

Four participants (P7, P9, P10, P12) employed visual metaphors to establish narrative coherence across frames. Unlike direct character or environmental consistency, these participants deliberately incorporated symbolic visual elements that conveyed conceptual meaning throughout the narrative. For example, during Phase 2, P7 explicitly requested recurring motifs: \textit{"In each image, I want a small clock somewhere in the scene, getting closer to midnight as the story progresses."} This metaphorical approach created a subtle visual thread connecting distinct scenes while symbolising the narrative's underlying theme of time pressure, she said:

\begin{quote}
\textit{"Please help me generate a 9-storyboard for a story. In each image, I want a small clock, gets closer to midnight as the story progresses. A mudered story happened between a monster and Ultraman. At 11:00 PM, in a peaceful small town with a clock on the town hall tower, Ultraman rests outside, and a monster lurks in the forest. By 11:15 PM, the monster attacks, knocking down buildings and scaring residents. At 11:20 PM, Ultraman awakens and rushes to the town. At 11:25 PM, they have their first confrontation, followed by a tough battle at 11:35 PM where the monster lands hits on Ultraman. By 11:45 PM, the monster has the upper hand, but Ultraman breaks free at 11:50 PM and devises a new strategy. At 11:55 PM, Ultraman unleashes his most powerful attack, and at midnight, he stands victorious as the monster is defeated and the townspeople cheer, before he flies away into the night."}
\end{quote}

\begin{figure*}
  \centering
  \includegraphics[width=\linewidth]{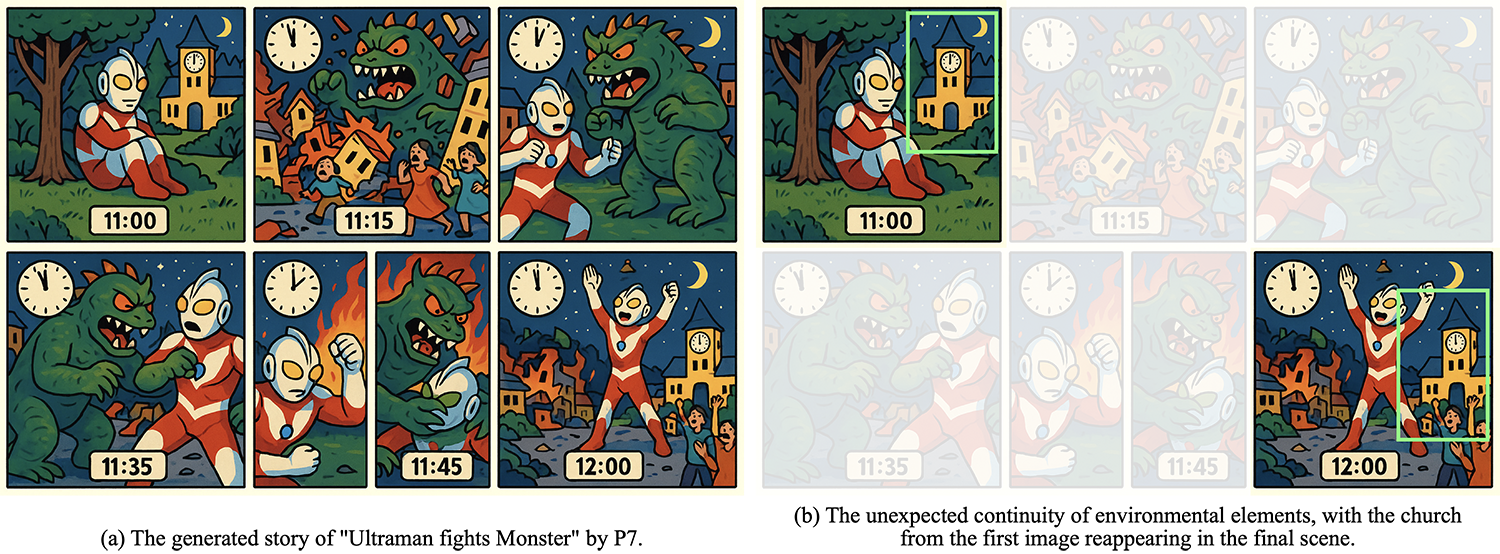}
  \caption{The generated story by P7, and our highlight of unexpected continuity.}
  \label{p7leftright}
\end{figure*}

The AI demonstrated mixed capability in maintaining these metaphorical elements. While it successfully implemented obvious visual metaphors (P7's clock appeared in each frame), it struggled with semantic relationships and narrative progression, and generated seven images instead of nine (see Fig.~\ref{p7leftright}a). For example, the clock requested by P7 appeared in the sky as an isolated element rather than being integrated meaningfully into the scene, suggesting the AI understood the literal instruction but not its narrative purpose. More concerning was the inconsistency in temporal progression. Although the storyline depicted Ultraman's battle lasting an hour, the clock time remained unchanged between the first and last images. Interestingly, environmental elements showed unexpected continuity, with the church from the first image reappearing in the final scene, creating an unintended cyclical narrative structure (see Fig.~\ref{p7leftright}b). Despite P7's attempts to show Ultraman gradually gaining the advantage, the AI persistently depicted the character at a disadvantage, failing to capture the intended narrative arc of overcoming challenges.

\begin{figure*}
  \centering
  \includegraphics[width=0.7\linewidth]{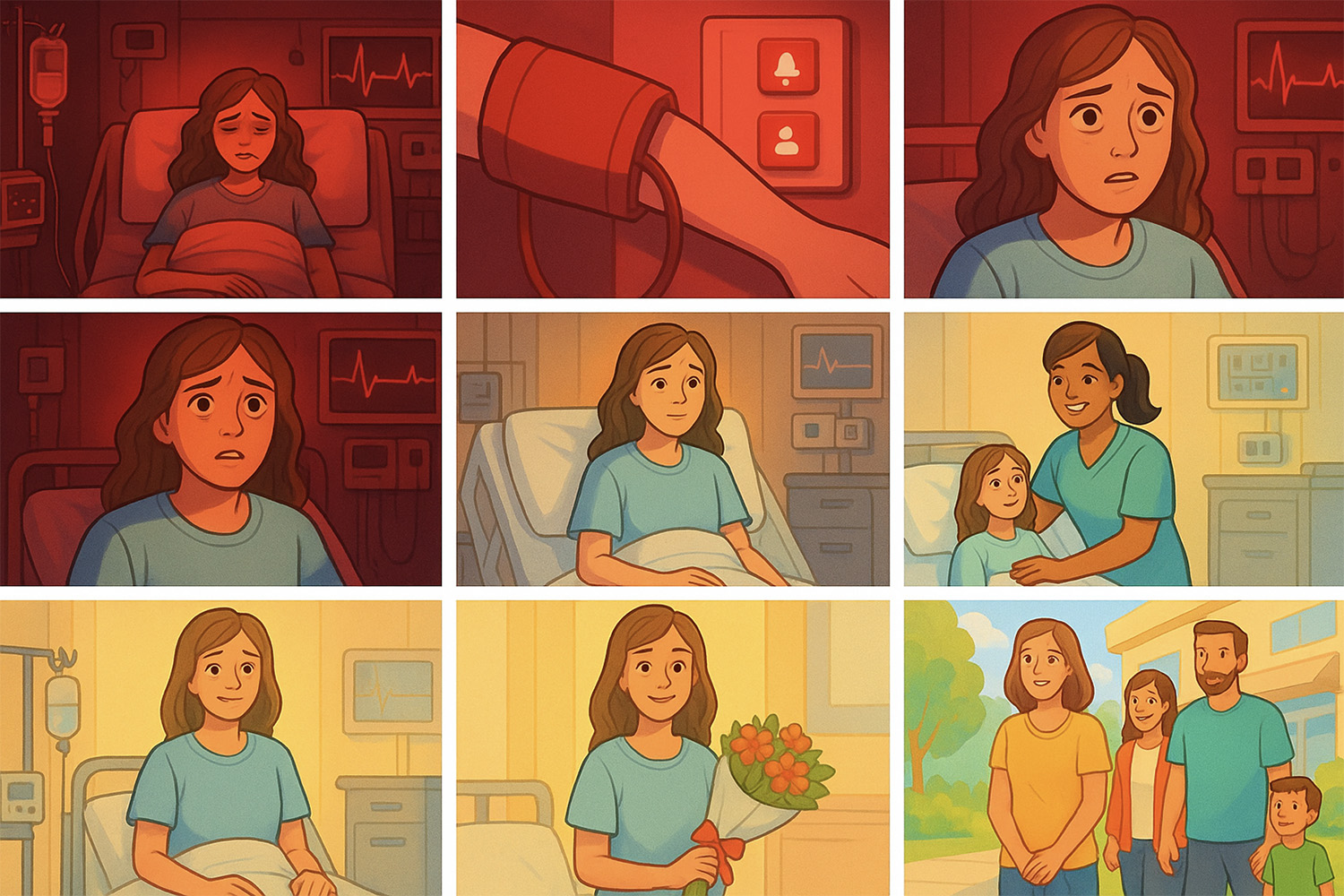}
  \caption{The generated story by P9.}
  \label{p9}
\end{figure*}

While P9 expressed frustration when attempting to implement color symbolism (see Fig.~\ref{p9}): \textit{"I wanted red elements to appear on different objects throughout the story, but the AI didn't understand this intention. Instead, it simply changed the background from red to yellow. The visual metaphor I tried to establish with the color red wasn't successfully represented."} This highlights a limitation in the AI's ability to grasp the semantic significance of visual elements beyond their literal representation. Her prompt is as follows:

\begin{quote}
\textit{"When Emma got sick, everything around her was red and scary. The hospital room had red lights, and the blood-pressure cuffs were bright red. Even the emergency call buttons on the wall were red. She felt so sick. But as she started getting better, the red surrounding her slowly faded. When she could finally sit up by herself, the red lights on the machines didn't seem as bright. When she was ready to go home, the only red thing left was a little red bow on the flowers her family had brought."}
\end{quote}

\subsubsection{\textbf{Strategy 2: Text-Image Complementarity}}
\begin{figure*}
  \centering
  \includegraphics[width=\linewidth]{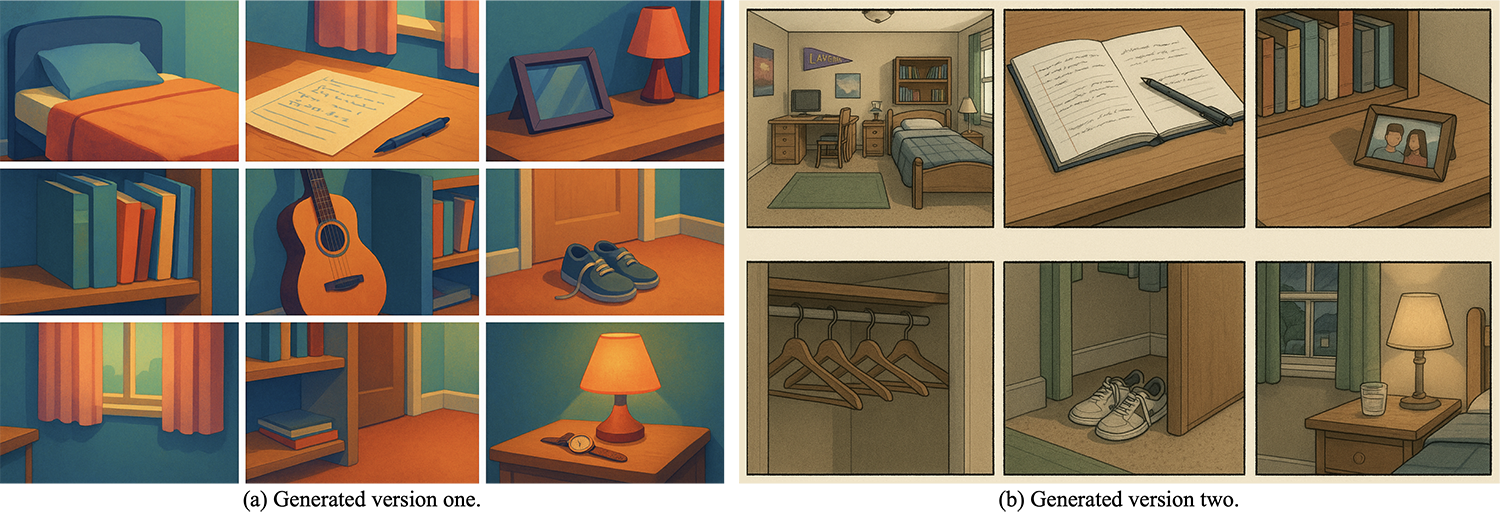}
  \caption{The generated stories by P11, show an empty bedroom of a teenager.}
  \label{p11}
\end{figure*}

Another particularly sophisticated intertextual strategy emerged among participants who deliberately designed complementary rather than redundant relationships between textual prompts and visual outputs. Rather than describing exactly what should appear visually, participants (P11, P13) crafted prompts that established what would remain unseen. For example (see Fig.~\ref{p11}), P11 offered: 

\begin{quote}
\textit{"Generate a storyboard of a teenager's bedroom that subtly shows someone is missing. One of the images should show an empty bedroom with subtle signs someone is missing, not showing the grief directly, but implying it through absence. The bed is neatly made but has a slight depression where someone recently sat. A half-finished homework assignment lies on the desk with a pen still uncapped. A framed photo of two friends is placed face down. Don't show any people in the storyboard."}
\end{quote}

P11 demonstrated this approach when creating a narrative about grief: \textit{"One of the images should show an empty bedroom with subtle signs someone is missing, not showing the grief directly, but implying it through absence."} This strategy created richer interpretive possibilities by establishing meaning in the gap between what was explicitly shown and what was textually suggested. During reflective interviews, P11 explained: \textit{"I found AI works better when I don't ask it to show everything in the text. When I leave space for it to visualize implications rather than statements, the results feel more emotionally authentic. This approach leaves more room for the audience's imagination, allowing them to participate in connecting elements of the story and imagine what kind of narrative this might be and what happened behind the scenes."}

The AI demonstrated strength under this kind of prompt, often generating more detailed visuals when responding to prompts focusing on single images rather than being direct in representing the whole. A clear distinction could be observed between the two versions produced by the AI in how it distributed P11's suggested content in generated two versions: version one emphasised close-up shots and narratives between these close perspectives (see Fig.~\ref{p11}a), while version two incorporated a long-distance shot in the first image combined with close-ups in the others (see Fig.~\ref{p11}b). This suggests that human-AI collaboration may achieve its intertextual results when humans manage the conceptual framing rather than through exhaustive description.

\subsection{Three Success Pathways (fsQCA)}

While the above analysis identifies individual factors influencing visual narrative success, the complex nature of human-AI collaboration suggests these factors likely work in combination rather than isolation. P4, P5, P7, P9, and P11 achieved similar levels of success despite demonstrating different factor combinations, suggesting multiple viable pathways to successful outcomes. This observation led us to investigate how these factors might systematically combine into distinct configuration patterns. Based on the qualitative findings presented in our study of human-AI collaborative narrative, we conducted fsQCA to systematically identify the configurations of conditions that lead to successful intertextual outcomes (see Appendix~\ref{appa}). Our necessity analysis presented no single condition that met the conventional threshold (consistency > 0.9) for being necessary for successful outcomes (see Table~\ref{tab:NecessityAnalysis}). Iteration Willingness (ITER) came closest with a consistency score of 0.87, suggesting that while not strictly necessary, a willingness to iterate was an important factor in almost all successful cases.

\begin{table}[h]
\centering
\caption{Necessity Analysis for Intertextual Success.}
\begin{tabular}{lcc}
\hline
Condition & Consistency & Coverage \\
\hline
COLLAB    & 0.76        & 0.82     \\
SCRIPT    & 0.82        & 0.78     \\
ITER      & 0.87        & 0.84     \\
TECH      & 0.65        & 0.74     \\
VISUAL    & 0.71        & 0.81     \\
EDUC      & 0.79        & 0.85     \\
\hline
\end{tabular}
\label{tab:NecessityAnalysis}
\end{table}

After applying a consistency threshold of 0.8 to filter configurations (following standard fsQCA procedures), our sufficiency analysis identified three distinct pathways to successful intertextual outcomes (see Table~\ref{tab:SufficiencyAnalysis}). These configurations show strong individual consistency, with pathway consistency scores ranging from 0.88 to 0.92.

\begin{table*}[h]
\centering
\caption{Sufficient Configurations for Intertextual Success.}
\begin{tabular}{lccc}
\hline
Configuration & Path Consistency & Path Coverage & Unique Coverage \\
\hline
COLLAB$\land$SCRIPT$\land$ITER$\land$EDUC & 0.92 & 0.45 & 0.28 \\
SCRIPT$\land$ITER$\land$TECH$\land$VISUAL & 0.91 & 0.39 & 0.22 \\
COLLAB$\land$ITER$\land$$\lnot$TECH$\land$VISUAL$\land$EDUC & 0.88 & 0.33 & 0.16 \\
\hline
\end{tabular}
\begin{flushleft}
\footnotesize{Note: $\land$ indicates the logical AND operator, $\lnot$ indicates the absence of a condition}
\end{flushleft}
\label{tab:SufficiencyAnalysis}
\end{table*}

\subsubsection{\textbf{Pathway 1: The Educational Collaborator Path \\(COLLAB$\land$SCRIPT$\land$ITER$\land$EDUC)}}

This configuration represents participants like P5 who viewed AI as a collaborative partner, explicitly separated narrative planning from visual execution, demonstrated willingness to iterate, and maintained an educational perspective throughout the process. These participants achieved success by approaching the collaboration as a learning experience, developing increasingly sophisticated prompting strategies through experimentation.

\subsubsection{\textbf{Pathway 2: The Technical Expert Path \\(SCRIPT$\land$ITER$\land$TECH$\land$VISUAL)}}

This path represents participants like P6 who combined technical knowledge about AI capabilities with strong visual literacy, methodically separating narrative development from visual execution and iterating as needed. These participants utilised their existing technical knowledge to work within the AI's strengths while mitigating its limitations.

\subsubsection{\textbf{Pathway 3: The Visual Thinker Path \\(COLLAB$\land$ITER$\land$$\lnot$TECH$\land$VISUAL$\land$EDUC)}}

This configuration represents participants like P11 who, despite limited technical knowledge ($\lnot$TECH), achieved success through strong visual thinking abilities, a collaborative mindset, willingness to iterate, and an educational perspective. These participants often relied on visual metaphors and developed what P8 referred to as "visual blueprints" to guide their collaboration.

Notably, no single condition is deemed necessary in the three identified pathways to successful human-AI collaboration in visual narrative. This absence implies the existence of multiple viable strategies for achieving successful outcomes. When comparing semi-open and fully open tasks, our fsQCA analysis also reveals that semi-open tasks favor the Educational Collaborator path, while fully open tasks exhibit stronger performance with the Technical Expert path, suggesting that different task constraints require distinct collaboration strategies. This analysis demonstrates how specific factor combinations lead to successful human-AI visual narrative outcomes, providing a deeper understanding than individual factors alone.

\section{Finding 3: Challenges and Desired Features (RQ3)}

We identified several consistent challenges that emerged across participants, highlighting the difficulties in human-AI co-creation that may benefit from further exploration and targeted interventions to improve collaboration and creative outcomes.

\subsection{Five Major Challenges}

\subsubsection{\textbf{Challenge 1: Navigating Cultural Understanding}}

Cultural understanding limitations emerged as a significant challenge in our study. As P5 explicitly noted that AI has difficulty with culturally specific interactions: \textit{"What it doesn't understand is the differences between countries. When I wrote instructions to it, I suddenly realised that it might not understand some of the cultural aspects in Chinese communication."} This cultural gap limited the AI's ability to represent authentic social dynamics in culturally specific contexts. P12 encountered similar challenges when attempting to depict traditional family gatherings, noting that the AI consistently produced Western-centric interpretations despite specific cultural references in the prompt. As he observed: \textit{"I tried to create a scene showing a Lunar New Year dinner, but the AI kept generating images that looked more like Christmas celebrations despite my detailed descriptions of specific foods and customs."} 

Also in P15's reflection, he elaborated on this limitation from a different angle, highlighting difficulties with culturally specific emotional expressions: \textit{"The subtleties of how emotions are displayed vary across cultures. When I tried to show restrained grief in an East Asian context, it kept generating overtly dramatic poses that felt culturally inappropriate."} These observations suggest that current image generation may require more culturally diverse training data and specialised prompting skills to accurately represent non-Western cultural contexts.

\subsubsection{\textbf{Challenge 2: Managing Iteration Process}}

Iteration management presented another common challenge, with participants (P4, P8) showing markedly different tolerance for refinement. While some embraced iteration (P5, P6, P15), others quickly abandoned tools after limited attempts. P4 explicitly stated this tendency: \textit{"Then I won't change, I can just not use it... Because if I've used this tool more than 3 times, I think it's not handy, I don't need to use it anymore."} 

\begin{figure*}
  \centering
  \includegraphics[width=\linewidth]{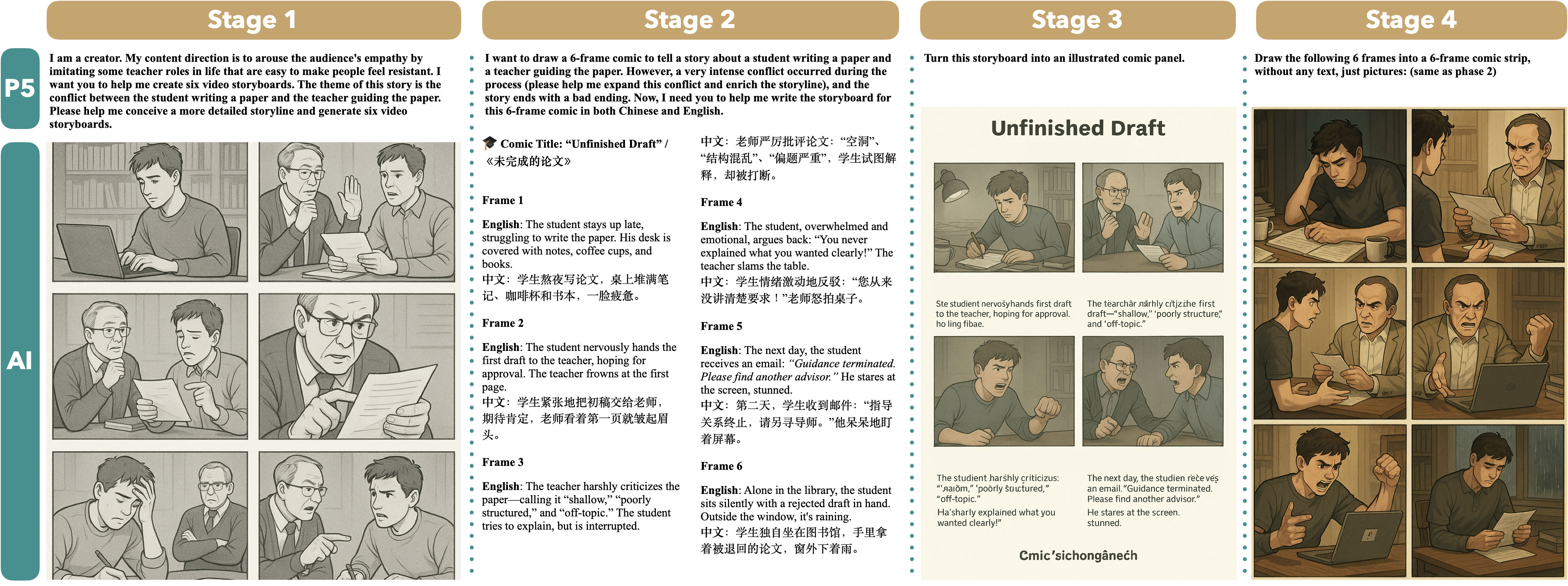}
  \caption{The P5's step-wise guidance.}
  \label{p52ai}
\end{figure*}

By contrast, P5's step-wise guidance philosophy, "\textit{still, guiding (AI) step by step}", progressively generated better results through targeted refinement in her open task. By establishing incremental approval processes and maintaining clear outcome expectations throughout, P5 effectively maintains creative control while leveraging AI capabilities, resulting in a coherent narrative progression from initial concept to finalised comic strip that efficiently addresses the student-teacher conflict theme (see Fig.~\ref{p52ai}). This balanced methodology provides scaffolded complexity that progresses from abstract concepts to concrete deliverables, accompanied by explicit bilingual content specifications and visual requirements at each stage.

\subsubsection{\textbf{Challenge 3: Maintaining Visual Consistency}}

\begin{figure*}
  \centering
  \includegraphics[width=\linewidth]{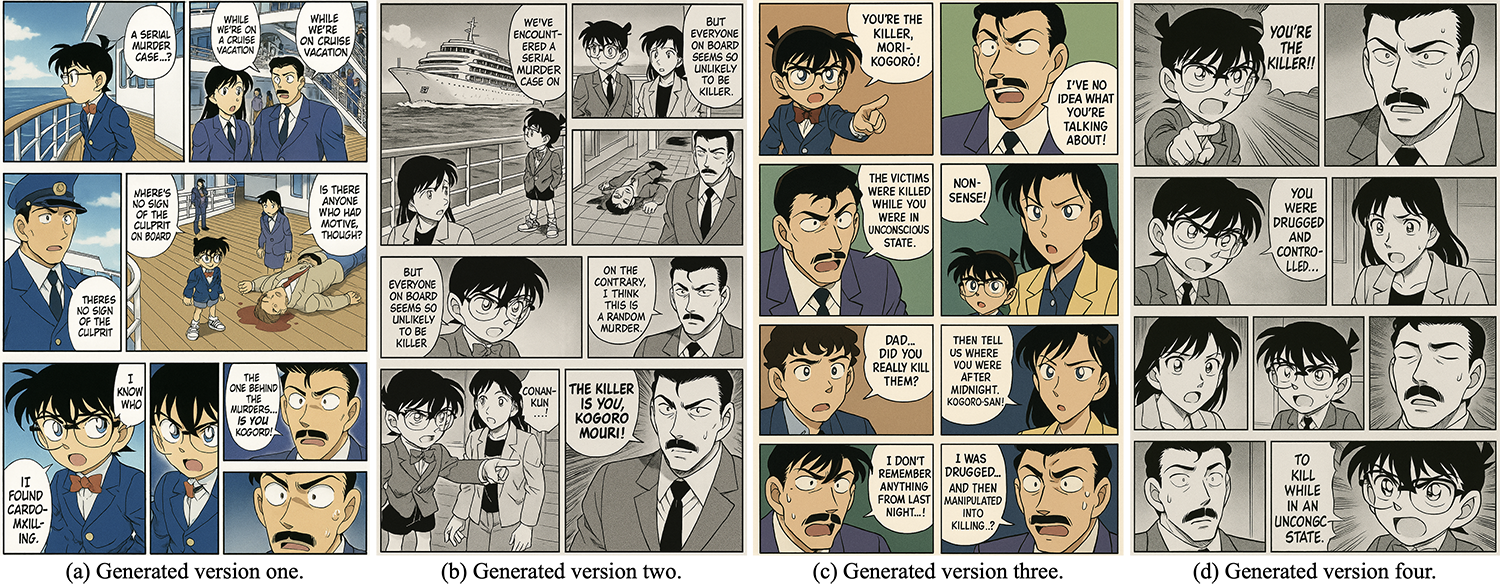}
  \caption{The generated stories by P6 in open task. In some images, Conan's height is the same as Mouri Ran's.}
  \label{p62}
\end{figure*}

Maintaining style and visual consistency across multiple images proved to be a persistent technical challenge. Character appearance, art style, and color schemes often varied unexpectedly between frames, as noted by P6 when examining inconsistencies in characters (see Fig.~\ref{p62}): "\textit{Yes, but in the first picture Conan is short.}" Some participants (P6, P10, P13) mitigated this issue by generating all panels simultaneously rather than sequentially, as P10 suggested: \textit{"Not generating them separately, but also generating them on one image, but not letting it generate six panels. What's the first panel? What's the second panel? Just send everything to it at once."}

\subsubsection{\textbf{Challenge 4: Translating Narrative to Visual Thinking}}

The most fundamental challenge was the translation between narrative and visual thinking. Converting linear narratives into discrete visual scenes proved particularly difficult for narrative-focused participants like P1, who often provided stream-of-consciousness descriptions that required substantial restructuring to function as visual prompts. The moderator had to intervene: "\textit{So, how do you paraphrase this sentence visually? How do you think you should paraphrase it?}" In contrast, participants who employed script-first approaches, like P4, P5, P7, and P8, more efficiently bridged this gap by explicitly separating narrative development from visual execution.

This difficulty reflects a deeper cognitive challenge in visual narrative: the shift from sequential, time-based thinking to spatial, moment-based representation. P7 articulated this struggle directly: \textit{"I know what will happen in my story, but deciding which specific moments to show and how to communicate the connections between them visually is much harder than I expected." The AI's literal interpretation tendencies further complicated this process, as it often failed to infer narrative connections humans take for granted, such as that the clock should be put on some foundations instead of on air."} And P8 developed a novel method that he called "visual blueprints", simplified sketches with annotations that separated narrative elements from visual direction: \textit{"I started thinking like a film director, distinguishing between what the audience needs to know and what they need to see, and making the text-image intertextual."} This approach forced explicit decisions about visual narrative principles before engaging with the AI.

\subsubsection{\textbf{Challenge 5: Representing Emotional Nuance}}

The representation of emotional nuance posed another significant challenge. Even participants who provided detailed character descriptions struggled to convey subtle emotional states through AI-generated imagery. Some participants (P2, P8, P13, P14) mentioned this challenge during reflective interviews. The think-aloud sessions showed that participants often resorted to explicit emotional labels rather than rich descriptions, with P2 simplifying complex emotions to basic terms like "happy," "sad," or "confused" despite initially attempting more sophisticated emotional portrayals. When he was reviewing the generated images, P2 noted: \textit{"I asked for 'wistful contemplation' but got something that looks more like generic sadness, AI seems to collapse emotional states into primary categories."}

The most successful emotional representation often came through environmental symbolism rather than facial expressions, as seen in P13's appreciation of how rain enhanced the emotional impact of a scene: \textit{"This rain, I didn't notice it at first, I only noticed the sad expression as he held the paper. Then, when you mentioned the rain, I realised this is a very good visual experience."} This environmental approach to emotion proved consistently more effective across multiple participants' projects, with P8 similarly using weather, lighting, and color palette to establish emotional tone rather than relying on character expressions.

This finding suggests a significant limitation in current AI systems' ability to render complex human expressions, particularly for emotions that exist on a spectrum rather than as discrete states. P14 directly addressed this limitation: \textit{"The AI seems to understand extreme emotions well, crying, laughing, rage, but struggles with the in-between states that make characters feel truly human."}

\subsection{Desired Role-Based Features}

We identified specific role-based features that would enhance human-AI collaborative visual storytelling: a continuity editor for character and environmental consistency, a layout composer for sequential arrangement and transitions, and a motif tracker for monitoring visual themes. Additional needs included a sequence-aware system that understands narrative context, a structure tool that links text annotations to visual elements, and bidirectional grounding enables dynamic connections where text changes inform visual generation and visual modifications suggest textual revisions.

\section{Discussion}

\subsection{Common User Workflow and AI Roles for Text-Image Intertextuality}

\begin{figure*}
  \centering
  \includegraphics[width=\linewidth]{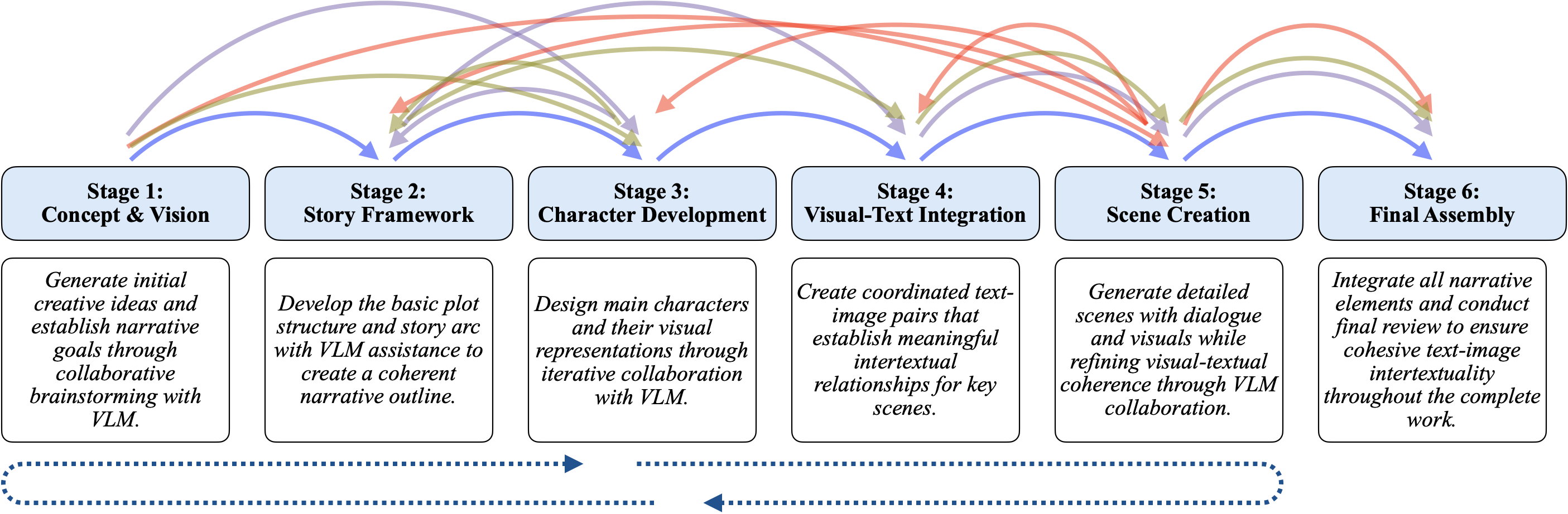}
  \caption{Common User Workflow with VLM (N=15). Stages 1-6 represent the main sequential workflow, while different colored arrows indicate varying user pathways based on individual preferences and collaborative approaches. Users may prioritise different stages or develop alternative sequences while adhering to the overall framework. The dotted arrow represents the overall cyclical nature of the collaborative creation process.}
  \label{commonworkflow}
\end{figure*}

\begin{figure*}
  \centering
  \includegraphics[width=\linewidth]{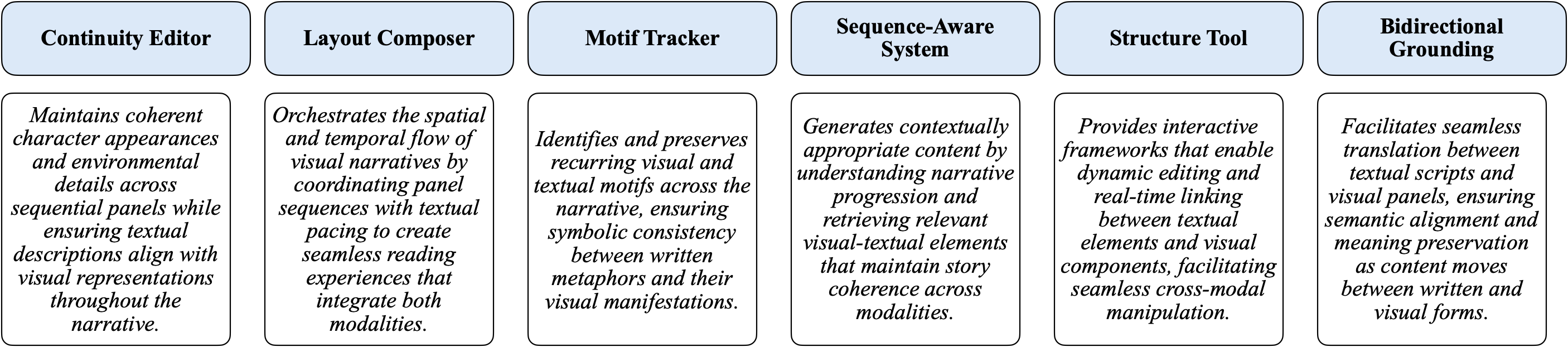}
  \caption{Summary of the six roles of AI in users' Expectations for facilitating \textit{text-image intertextuality}.}
  \label{roles}
\end{figure*}

Anchored in Fig.~\ref{commonworkflow}, the cyclical six-stage journey indicates that success hinges on treating AI as the six-role co-developmental partners (see Fig.~\ref{roles}): continuity editor, layout composer, motif tracker, sequence-aware system, structure tool, and bidirectional grounding. Participants who achieved the most successful outcomes (P4, P5, P7, P9, and P11) transcended tool-centric or assistant paradigms, iteratively co-evolving both narrative and their own creative competence. P5 explicitly credits AI for sharpening her expressive skills~\cite{rezwana2022understanding,rezwana2023designing,mahmud2023study}. This mutual pedagogical growth closes the "huge gap" (P11) between narrative intent and visual execution. Novices invent "visual blueprints" (P8) and other prompt-craft tactics, converting model limitations into scaffolded lessons on composition, metaphor, sequence, and positioning, thereby generating systems as implicit co-educators rather than mere executors. The pivotal move is to foster intertextual "braiding"~\cite{groensteen2007system}: instead of dictating every pixel, authors establish conceptual frames and emotional tones, "leaving space for AI to visualise implications rather than statements" (P11), thereby inviting viewers to co-create meaning across the semantic gap between text and image.

\subsection{Future Directions and Design Implications}

\begin{figure*}
  \centering
  \includegraphics[width=\linewidth]{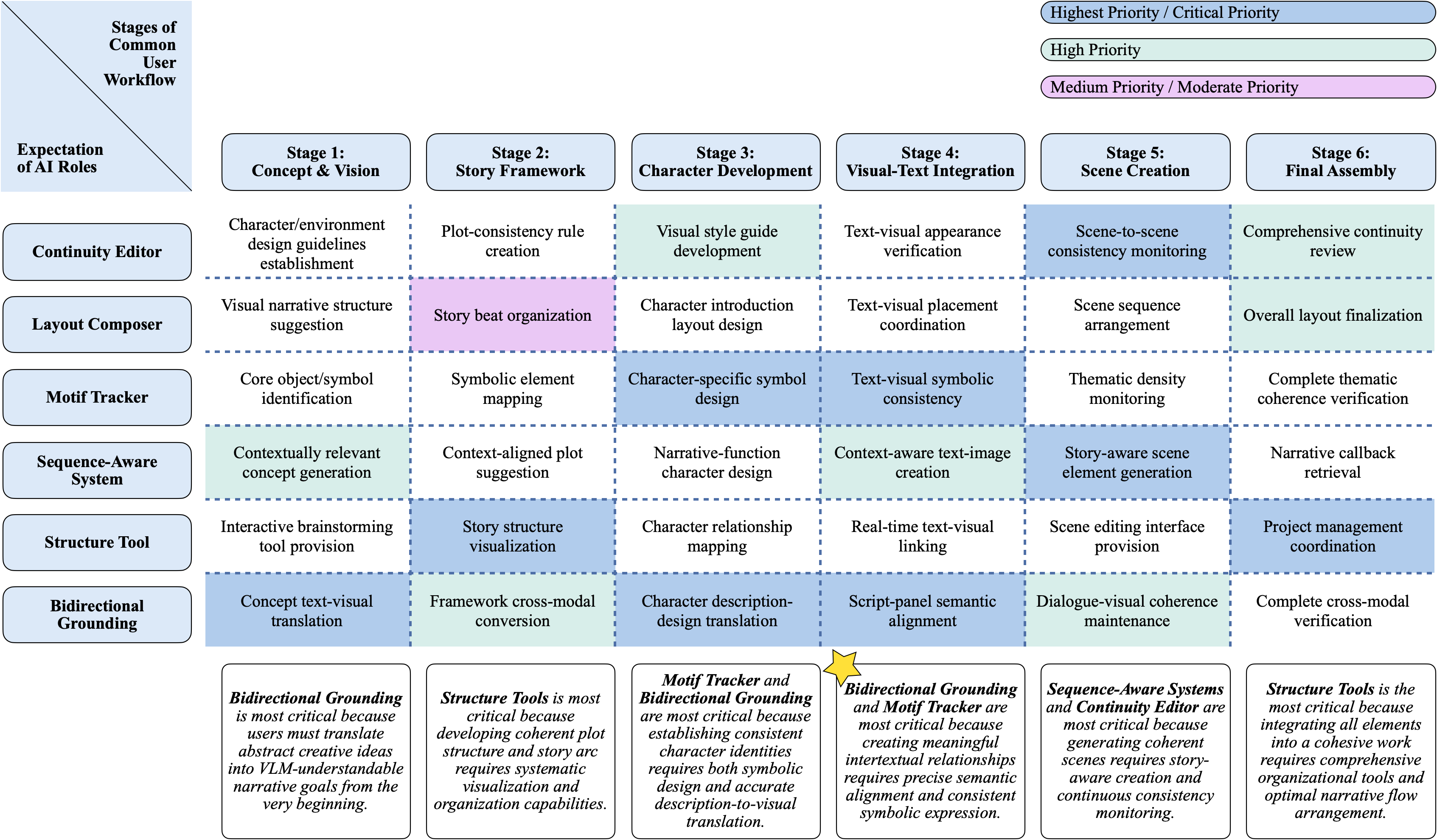}
  \caption{Stage 4 serves as the critical turning point for establishing intertextuality, with Bidirectional Grounding and Motif Tracker being most essential. Stage 5 represents the intensive phase for maintaining coherence, where Sequence-Aware Systems and Continuity Editor bear the greatest burden. Each stage has its unique core challenges requiring specialised support from different roles.}
  \label{priority}
\end{figure*}

Fig.~\ref{priority} illustrates critical intervention points where future VLM systems could deploy specialised AI roles. At Stage 4's pivotal moment, Bidirectional Grounding and Motif Tracker capabilities become essential for stabilizing text-image intertextuality, enabling users to establish meaningful semantic relationships rather than literal translations. During Stage 5's intensive coherence phase, Sequence-Aware Systems and Continuity Editor functions need to maintain character consistency, environmental continuity, and visual metaphor alignment across sequential panels. The study's finding that multi-stage workflows significantly increased successful intertextuality reinforces the need for stage-specific interventions that adapt dynamically to users' current workflow position, providing targeted support when collaborative challenges peak.

\begin{framed}
\noindent
\textbf{\large Design Implication 1: Implementing Workflow-Aware AI Assistance} \vspace{1mm}

Future VLM interfaces could implement context-sensitive AI roles~\cite{hu2025vision} that activate based on the user's current stage in the creative process. Given that participants who separated narrative development from visual execution achieved more coherent results, AI could automatically detect workflow transitions and surface appropriate tools without overwhelming users with irrelevant features. This design will encourage the successful multi-stage approach observed in participants like P13, he noted: \textit{"After revisions, once I had the script, I understood the plot better, and then bringing that into this, I felt the script became very complete"}, to ensure that AI support matches the cognitive demands of each creative phase.

\end{framed}

The study represents how novices \textbf{learn through productive} friction with AI capabilities, transforming prompt failures into creative breakthroughs. Participants who adopted an "educational perspective" showed greater iteration willingness and higher satisfaction, viewing challenges as learning opportunities rather than failures. Rather than eliminating all user challenges, interfaces could embed just-in-time scaffolding of visual literacy principles, helping users understand compositional choices, narrative structure, and the implementation of metaphors. This approach requires designing for interpretability and iterative feedback, enabling users to comprehend how their creative decisions translate into visual outcomes while building long-term creative agency.

\begin{framed}
\noindent
\textbf{\large Design Implication 2: Fostering Educational Mindsets Through Interface Design} \vspace{1mm}

Next-generation systems could actively promote the educational perspective that correlates with success by reframing interactions as learning opportunities. When users encounter prompt failures or unexpected outputs, interfaces should offer contextual explanations about visual storytelling techniques~\cite{belz2024story}, cultural symbolism~\cite{hamna2025kahani}, or compositional theory. Systems may include progress tracking and reflection prompts that help users recognise their skill development, transforming AI "mistakes" into teachable moments while maintaining the growth mindset that enabled participants like P5 to view AI collaboration as \textit{"... now I see AI as a process of exercising myself; using it can actually train my expression very well. Or some descriptions of things. My feeling is that this is also a new learning process."} 

\end{framed}

Participants' relationship with AI strongly influenced success rates, with those viewing AI as a collaborative partner achieving higher success than those treating it purely as a tool. The contrast between P5's collaborative approach (\textit{"I feel it's more like collaboration, because sometimes it's feedback tells me that if I completely command it without considering its abstract thinking, what it gives me rarely meets my needs."}) and P11's command-based frustration (\textit{"I feel more like I'm commanding it to do things... because I'm not very satisfied with what it generates"}) demonstrates how relationship framing affects outcomes. Future systems must foster collaborative rather than hierarchical human-AI relationships~\cite{fragiadakis2024evaluating} by providing feedback mechanisms that highlight AI's reasoning process and encourage user adaptation based on AI capabilities and limitations.

\begin{framed}
\noindent
\textbf{\large Design Implication 3: Designing for Collaborative Partnership Recognition} \vspace{1mm}

AI Interfaces could explicitly frame interactions as partnerships through language choices, feedback mechanisms, and collaborative features~\cite{edwards2025human}. These systems should provide transparency into AI decision-making processes~\cite{pillai2024enhancing,buijsman2024transparency}, enabling users to understand how their inputs influence outputs and fostering the mutual adaptation that characterizes successful partnerships. This includes implementing conversational interfaces that acknowledge user input while explaining AI reasoning, moving beyond simple command-response patterns~\cite{figueiredo2025designing} toward genuine collaborative dialogue.

\end{framed}

Participants with comparative knowledge of AI systems demonstrated more strategic tool utilisation, working within the tool's capabilities rather than against them. The inverse relationship between required assistance and intertextual success suggests that baseline technical proficiency has a significant impact on collaborative outcomes. Systems could address this disparity by providing adaptive support that builds user competency without creating dependency, ensuring that technical knowledge becomes a pathway to creative agency rather than a barrier to access.

\begin{framed}
\noindent
\textbf{\large Design Implication 4: Adaptive Competency Building} \vspace{1mm}

Future VLM systems could implement progressive competency scaffolding that adapts to individual user skill levels while building toward technical autonomy. For novice users, systems can provide guided workflows and contextual explanations that build understanding of AI capabilities and limitations~\cite{mathew2025recent}. For advanced users, systems may offer expanded control options and comparative tool information. This adaptive approach will monitor user progress and gradually reduce assistance as competency develops, ensuring that all users can eventually achieve the strategic tool utilisation demonstrated by technically proficient participants.

\end{framed}

To support the development of coherent visual narratives, AI systems could maintain a persistent memory of prior images~\cite{martin2025ai}, character designs, and recurring motifs throughout extended creative sessions. This architectural requirement addresses consistency challenges that participants frequently encountered, particularly in character appearance and environmental details across panels. Additionally, systems can incorporate culturally diverse training data and specialised prompting interfaces that recognize cultural specificity in emotional expression, social dynamics, and visual symbolism.

\begin{framed}
\noindent
\textbf{\large Design Implication 5: Building Cultural Memory Systems} \vspace{1mm}

Future VLM architectures could implement dual-layer memory systems: technical memory for visual consistency (character features, color palettes, environmental details) and cultural memory for contextually appropriate representation~\cite{debnath2025lightstate}. These systems allow users to specify cultural contexts and automatically suggest culturally relevant visual elements, addressing representation gaps while maintaining narrative coherence across sequential panels. The correlation between workflow structure and visual coherence suggests that these memory systems will integrate with stage-aware interfaces to provide continuity support when users transition between creative phases.

\end{framed}

\section{Conclusion}

This study examined how novice users collaborate with VLMs to create visual narratives, with a specific focus on the emergence and negotiation of \textit{text-image intertextuality}, the dynamic process of meaning-making between textual prompts and AI-generated visuals. Through a three-phase qualitative study with 15 participants using GPT-4o, we investigated how users allocate tasks, establish intertextual connections, and navigate challenges in co-creating sequential visual stories. Our findings show that participants learned to utilize AI's "semantic surplus," employing visual augmentations that extended beyond literal prompt descriptions to enrich narrative meaning. Our fsQCA analysis presents us with three distinct pathways to successful intertextual collaboration: the \textit{Educational Collaborator}, the \textit{Technical Expert}, and the \textit{Visual Thinker} paths, where willingness to iterate plays a key factor across all of them, and identified participants' adaptive and reflective engagement with AI tools. However, they also faced significant challenges, including cultural representation gaps, visual inconsistency, and difficulties in translating narrative intent into visual form. Our work contributes to HCI and AI-mediated creativity research by providing an empirical account of how non-experts negotiate meaning with VLMs, identifying key strategies and challenges in establishing text-image intertextuality, and proposing design implications for role-based AI assistants that support iterative, human-led creative processes, while highlighting the pedagogical potential where friction and failure become opportunities for learning and creative growth that should inform future development of more interpretable, sequence-aware AI systems.

\begin{acks}

\end{acks}

\bibliographystyle{ACM-Reference-Format}
\bibliography{main}

\appendix

\section{fsQCA Methodology and Calibration Process}~\label{appa}

We performed our analysis using fsQCA 4.1 Mac\footnote{https://sites.socsci.uci.edu/~cragin/fsQCA/software.shtml} software following established procedures.

\subsection{Calibration Standards and Process}
For calibrating our six condition variables and outcome variable into fuzzy-set membership scores (0.0-1.0), we established qualitative anchors defining full membership (1.0), the crossover point (0.5), and full non-membership (0.0) for each variable:

\begin{itemize}
    \item \textbf{Collaborative Mindset (COLLAB)}:
    \begin{enumerate}
        \item Full membership (1.0): Explicitly expressed view of AI as a partner, as exemplified by P5: \textit{"its feedback helps me learn how to better work with it"}
        \item Crossover point (0.5): Mixed perspective, sometimes commanding, sometimes collaborating
        \item Full non-membership (0.0): Purely instrumental view, as shown by P4: "\textit{after using it three times without effect, there's no need to use it anymore}"
    \end{enumerate}
    
    \item \textbf{Script-First Approach (SCRIPT)}:
    \begin{enumerate}
        \item Full membership (1.0): Complete separation of narrative development from visual execution
        \item Crossover point (0.5): Partial separation, sometimes mixed development
        \item Full non-membership (0.0): Simultaneous development of narrative and visual elements
    \end{enumerate}
    
    \item \textbf{Iteration Willingness (ITER)}:
    \begin{enumerate}
        \item Full membership (1.0): Sustained prompt modification across multiple attempts ($\geq$3 iterations)
        \item Crossover point (0.5): Moderate level of iteration (1-2 iterations)
        \item Full non-membership (0.0): Abandonment after initial attempts
    \end{enumerate}
    
    \item \textbf{Technical Knowledge (TECH)}:
    \begin{enumerate}
        \item Full membership (1.0): Demonstrated sophisticated understanding of multiple AI platforms, as shown by P6: \textit{"I think ChatGPT is slightly weaker in image generation"}
        \item Crossover point (0.5): Basic understanding of AI capabilities and limitations
        \item Full non-membership (0.0): No prior experience, requiring consistent guidance
    \end{enumerate}
    
    \item \textbf{Visual Literacy (VISUAL)}:
    \begin{enumerate}
        \item Full membership (1.0): Fluent translation of narrative concepts to visual elements, rich visual vocabulary
        \item Crossover point (0.5): Basic description of visual elements, lacking detail
        \item Full non-membership (0.0): Purely textual thinking, inability to effectively communicate visual concepts
    \end{enumerate}
    
    \item \textbf{Educational Perspective (EDUC)}:
    \begin{enumerate}
        \item Full membership (1.0): Explicit learning orientation, as exemplified by P12: \textit{"using AI is actually a process for exercising my own expression abilities"}
        \item Crossover point (0.5): Partial recognition of learning opportunities
        \item Full non-membership (0.0): Purely outcome-focused, no growth mindset
    \end{enumerate}
\end{itemize}

For the outcome variable \textbf{Intertextual Success (SUCCESS)}, we evaluated participants' final visual narratives based on the coherence, the effective establishment of text-image connections, and the narrative quality. During this process, we invited three professionals with a strong background in formal aesthetics, visual analysis, and practical skills to support us in considering the visual outcomes:

\begin{itemize}
    \item Full membership (1.0): Sophisticated intertextual relationships, clear narrative flow, high visual coherence
    \item Crossover point (0.5): Basic intertextual relationships, moderate coherence
    \item Full non-membership (0.0): Failure to establish connections between images and text
\end{itemize}

The calibration process involved two researchers independently reviewing all data sources (interview transcripts, think-aloud sessions, video observation notes) and scoring each participant on each condition according to the above standards. For cases with scoring discrepancies, researchers discussed the issue until a consensus was reached.

\subsection{Analysis Procedure}
Our fsQCA analysis followed these steps:
\begin{enumerate}
    \item Construction of a truth table showing all possible configurations of conditions
    \item Application of frequency and consistency thresholds (0.8) to filter configurations
    \item Logical minimisation using the Quine-McCluskey algorithm
    \item Analysis of complex, parsimonious, and intermediate solutions
\end{enumerate}

We selected the intermediate solution for our final report, which incorporates theoretical knowledge about how conditions should relate to the outcome. For assessing the quality of necessity and sufficiency relations, we calculated consistency and coverage measures following standard fsQCA practices:

\begin{itemize}

  \item \textbf{Necessity Consistency}:
    \begin{equation}
      \mathrm{Consistency}(X \le Y)
        = \frac{\sum \min(X_i, Y_i)}{\sum X_i}
    \end{equation}

  \item \textbf{Necessity Coverage}:
    \begin{equation}
      \mathrm{Coverage}(X \le Y)
        = \frac{\sum \min(X_i, Y_i)}{\sum Y_i}
    \end{equation}

  \item \textbf{Path Consistency}:
    \begin{equation}
      \mathrm{Consistency}(X \le Y)
        = \frac{\sum \min(X_i, Y_i)}{\sum X_i}
    \end{equation}

  \item \textbf{Path Coverage}:
    \begin{equation}
      \mathrm{Coverage}(X \le Y)
        = \frac{\sum \min(X_i, Y_i)}{\sum Y_i}
    \end{equation}

  \item \textbf{Unique Coverage}:
    \begin{equation}
      \mathrm{UniCoverage}
        = \mathrm{PathCoverage}
        - \mathrm{Coverage}(\text{shared solution})
    \end{equation}

\end{itemize}
\end{document}